\documentclass[twocolumn,reprint,prc,superscriptaddress,amsfonts,amssymb,amsmath,showpacs]{revtex4-1}

\usepackage{longtable}
\usepackage{epsfig}
\usepackage{graphicx}
\usepackage{dcolumn}
\usepackage{bm}

\begin{document}

\title{Measurement of the normalized $^{238}$U(n,f)/$^{235}$U(n,f) cross section ratio\\from threshold to 30 MeV with the fission Time Projection Chamber}

\author{R.~J.~Casperson}\email{casperson1@llnl.gov}\affiliation{Lawrence Livermore National Laboratory, Livermore, California 94550}
\author{D.~M.~Asner}\affiliation{Pacific Northwest National Laboratory, Richland, Washington 99354}
\author{J.~Baker}\thanks{deceased}\affiliation{Idaho National Laboratory, Idaho Falls, Idaho 83402}
\author{R.~G.~Baker}\affiliation{California Polytechnic State University, San Luis Obispo, California 93407}
\author{J.~S.~Barrett}\affiliation{Oregon State University, Corvallis, Oregon 97331}
\author{N.~S.~Bowden}\affiliation{Lawrence Livermore National Laboratory, Livermore, California 94550}
\author{C.~Brune}\affiliation{Ohio University, Athens, Ohio 45701}
\author{J.~Bundgaard}\affiliation{Colorado School of Mines, Golden, Colorado 80401}
\author{E.~Burgett}\affiliation{Idaho State University, Pocatello, Idaho 83209}
\author{D.~A.~Cebra}\affiliation{University of California, Davis, California 95616}
\author{T.~Classen}\affiliation{Lawrence Livermore National Laboratory, Livermore, California 94550}
\author{M.~Cunningham}\affiliation{Lawrence Livermore National Laboratory, Livermore, California 94550}
\author{J.~Deaven}\affiliation{Idaho State University, Pocatello, Idaho 83209}
\author{D.~L.~Duke}\affiliation{Colorado School of Mines, Golden, Colorado 80401}\affiliation{Los Alamos National Laboratory, Los Alamos, New Mexico 87545}
\author{I.~Ferguson}\affiliation{Georgia Institute of Technology, Atlanta, Georgia 30332}
\author{J.~Gearhart}\affiliation{University of California, Davis, California 95616}
\author{V.~Geppert-Kleinrath}\affiliation{Los Alamos National Laboratory, Los Alamos, New Mexico 87545}
\author{U.~Greife}\affiliation{Colorado School of Mines, Golden, Colorado 80401}
\author{S.~Grimes}\affiliation{Ohio University, Athens, Ohio 45701}
\author{E.~Guardincerri}\affiliation{Los Alamos National Laboratory, Los Alamos, New Mexico 87545}
\author{U.~Hager}\affiliation{Colorado School of Mines, Golden, Colorado 80401}
\author{C.~Hagmann}\affiliation{Lawrence Livermore National Laboratory, Livermore, California 94550}
\author{M.~Heffner}\affiliation{Lawrence Livermore National Laboratory, Livermore, California 94550}
\author{D.~Hensle}\affiliation{Colorado School of Mines, Golden, Colorado 80401}
\author{N.~Hertel}\affiliation{Georgia Institute of Technology, Atlanta, Georgia 30332}
\author{D.~Higgins}\affiliation{Colorado School of Mines, Golden, Colorado 80401}
\author{T.~Hill}\affiliation{Idaho State University, Pocatello, Idaho 83209}
\author{L.~D.~Isenhower}\affiliation{Abilene Christian University, Abilene, Texas 79699}
\author{J.~King}\affiliation{Oregon State University, Corvallis, Oregon 97331}
\author{J.~L.~Klay}\affiliation{California Polytechnic State University, San Luis Obispo, California 93407}
\author{N.~Kornilov}\affiliation{Ohio University, Athens, Ohio 45701}
\author{R.~Kudo}\affiliation{California Polytechnic State University, San Luis Obispo, California 93407}
\author{A.~B.~Laptev}\affiliation{Los Alamos National Laboratory, Los Alamos, New Mexico 87545}
\author{W.~Loveland}\affiliation{Oregon State University, Corvallis, Oregon 97331}
\author{M.~Lynch}\affiliation{California Polytechnic State University, San Luis Obispo, California 93407}
\author{W.~S.~Lynn}\affiliation{Abilene Christian University, Abilene, Texas 79699}
\author{J.~A.~Magee}\affiliation{Lawrence Livermore National Laboratory, Livermore, California 94550}
\author{B.~Manning}\affiliation{Los Alamos National Laboratory, Los Alamos, New Mexico 87545}
\author{T.~N.~Massey}\affiliation{Ohio University, Athens, Ohio 45701}
\author{C.~McGrath}\affiliation{Idaho State University, Pocatello, Idaho 83209}
\author{R.~Meharchand}\affiliation{Los Alamos National Laboratory, Los Alamos, New Mexico 87545}
\author{M.~P.~Mendenhall}\affiliation{Lawrence Livermore National Laboratory, Livermore, California 94550}
\author{L.~Montoya}\affiliation{Los Alamos National Laboratory, Los Alamos, New Mexico 87545}
\author{N.~T.~Pickle}\affiliation{Abilene Christian University, Abilene, Texas 79699}
\author{H.~Qu}\affiliation{Abilene Christian University, Abilene, Texas 79699}
\author{J.~Ruz}\affiliation{Lawrence Livermore National Laboratory, Livermore, California 94550}
\author{S.~Sangiorgio}\affiliation{Lawrence Livermore National Laboratory, Livermore, California 94550}
\author{K.~T.~Schmitt}\affiliation{Los Alamos National Laboratory, Los Alamos, New Mexico 87545}
\author{B.~Seilhan}\affiliation{Lawrence Livermore National Laboratory, Livermore, California 94550}
\author{S.~Sharma}\affiliation{Abilene Christian University, Abilene, Texas 79699}
\author{L.~Snyder}\affiliation{Lawrence Livermore National Laboratory, Livermore, California 94550}
\author{S.~Stave}\affiliation{Pacific Northwest National Laboratory, Richland, Washington 99354}
\author{A.~C.~Tate}\affiliation{Abilene Christian University, Abilene, Texas 79699}
\author{G.~Tatishvili}\affiliation{Pacific Northwest National Laboratory, Richland, Washington 99354}
\author{R.T.~Thornton}\affiliation{Abilene Christian University, Abilene, Texas 79699}
\author{F.~Tovesson}\affiliation{Los Alamos National Laboratory, Los Alamos, New Mexico 87545}
\author{D.~E.~Towell}\affiliation{Abilene Christian University, Abilene, Texas 79699}
\author{R.~S.~Towell}\affiliation{Abilene Christian University, Abilene, Texas 79699}
\author{N.~Walsh}\affiliation{Lawrence Livermore National Laboratory, Livermore, California 94550}
\author{S.~Watson}\affiliation{Abilene Christian University, Abilene, Texas 79699}
\author{B.~Wendt}\affiliation{Idaho State University, Pocatello, Idaho 83209}
\author{L.~Wood}\affiliation{Pacific Northwest National Laboratory, Richland, Washington 99354}
\author{L.~Yao}\affiliation{Oregon State University, Corvallis, Oregon 97331}
\author{W.~Younes}\affiliation{Lawrence Livermore National Laboratory, Livermore, California 94550}

\collaboration{NIFFTE Collaboration}\homepage{http://niffte.calpoly.edu/}\noaffiliation

\date{\today}

\begin{abstract}
The normalized $^{238}$U(n,f)/$^{235}$U(n,f) cross section ratio has been measured using the NIFFTE fission Time Projection Chamber from the reaction threshold to $30$~MeV.  The fissionTPC is a two-volume MICROMEGAS time projection chamber that allows for full three-dimensional reconstruction of fission-fragment ionization profiles from neutron-induced fission.
The measurement was performed at the Los Alamos Neutron Science Center, where the neutron energy is determined from neutron time-of-flight.  The $^{238}$U(n,f)/$^{235}$U(n,f) ratio reported here is the first cross section measurement made with the fissionTPC, and will provide new experimental data for evaluation of the $^{238}$U(n,f) cross section, an important standard used in neutron-flux measurements.  
Use of a development target in this work prevented the determination of an absolute normalization, to be addressed in future measurements.  Instead, the measured cross section ratio has been normalized to ENDF/B-VIII.$\beta$5 at 14.5 MeV.
\end{abstract}


\maketitle

\section{\label{sec:Introduction}Introduction}

Neutron-induced reactions play an essential role in nucleosynthesis,  advanced nuclear reactor design, and stockpile stewardship.  Future nuclear reactors may use fast neutrons, which would allow for more energy to be extracted from the fuel, and reduce the lifetime of nuclear waste.  Since the neutron spectrum in fast reactors is different than in light water reactors, greater precision neutron-induced cross section data at higher energies is required~\cite{Aliberti2006ANE,Aliberti2008NDS,Salvatores2008NDS}.
Cross section data is typically fit with nuclear reaction theory models such as EMPIRE~\cite{Herman2007NDS}, TALYS~\cite{Koning2007NDST}, and GNASH~\cite{Young1997LA}, which allow for determination of quantities such as fission barrier heights, nuclear level densities, and fission fragment anisotropy in the compound nucleus.  The normalized cross section ratio measured in this work can be used to better understand these nuclear properties.

The fission Time Projection Chamber (fissionTPC) is a two-volume MICROMEGAS detector designed and built by the NIFFTE (Neutron-Induced Fission Fragment Tracking Experiment) Collaboration to measure neutron-induced fission cross sections with high precision~\cite{Heffner2014NIMA}.  Ionization tracks deposited by fission fragments, $\alpha$-particles, and recoils from neutron-scattering are drifted across each chamber, and an electron avalanche multiplies the charge before collection on a pixelated pad plane.  Full three-dimensional track reconstruction is used for particle identification and determination of the fission fragment detection efficiency and related systematic uncertainties.

Past neutron-induced fission cross section measurements have used parallel-plate ionization chambers~\cite{Wender1993NIMA}, which include stacks of
foils separated by a distance smaller than the typical particle range.  Light ions such as $\alpha$-particles have a longer range and much smaller stopping power than fission fragments, and deposit very little energy in the space between foils.  Fission fragments have much higher stopping power between the foils, and can usually be distinguished from $\alpha$-particles.  Twin Frisch-grid ionization chambers~\cite{Salvador2015PRC1,Salvador2015PRC2} allow for an inference of the charged-particle track angle, which provides additional information for determination of the fission fragment detection efficiency.  
The fissionTPC has the additional capability of full three-dimensional track reconstruction, which provides the particle's origin, energy, angle, length, and ionization profile. In addition to providing more information from
which to determine detection efficiency, these quantities also allow in situ measurement of the target atom density and neutron beam flux. 

\begin{figure*}[t!]
\includegraphics[scale=0.9]{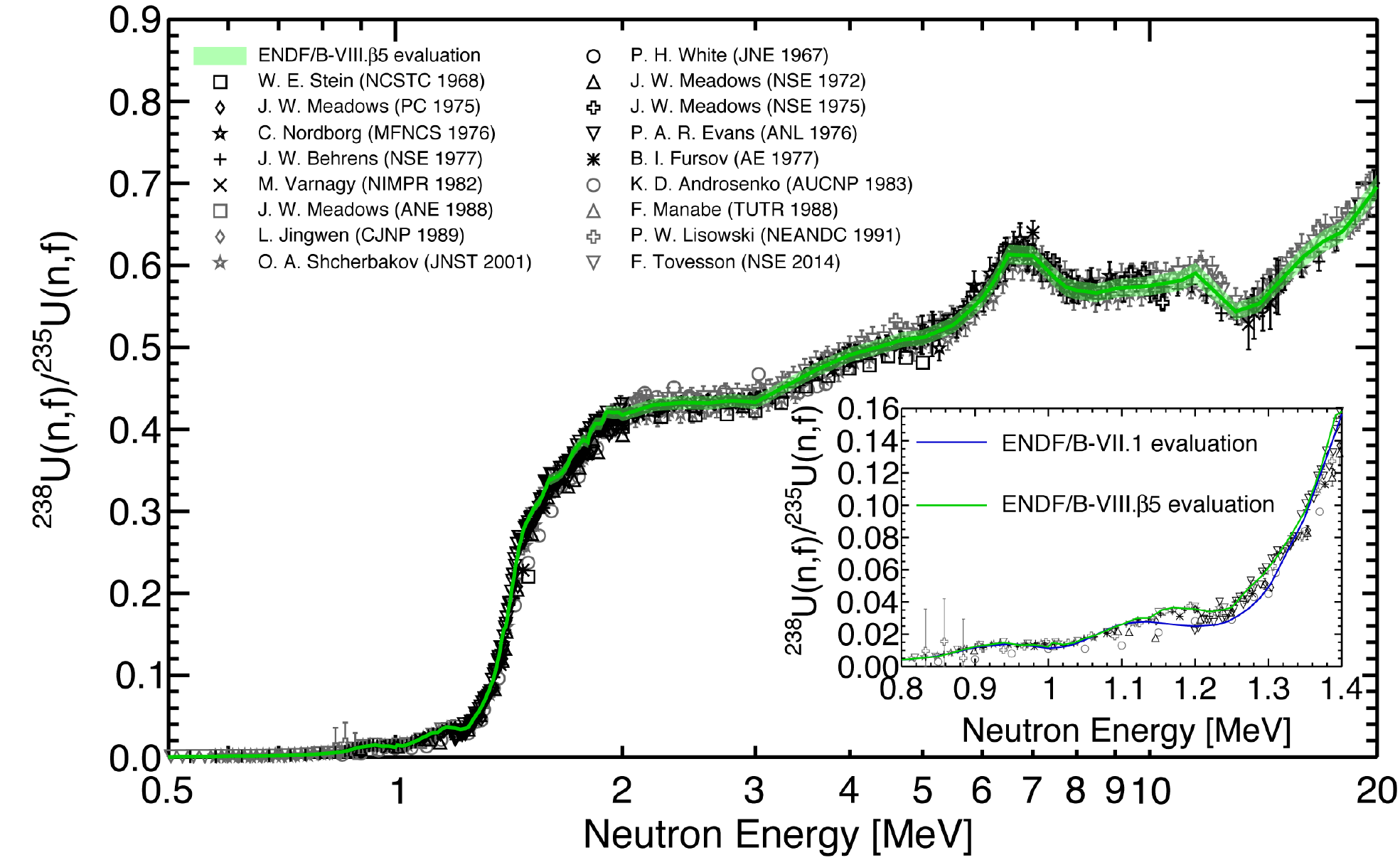}
\caption{\label{fig:FullU8U5Ratio} (Color online) Past measurements of the $^{238}$U(n,f)/$^{235}$U(n,f) cross section ratio shown between $0.5 - 30$~MeV~\cite{White1967JNE,Stein1968CP,Meadows1972NSE,Meadows1975NSE,Meadows1975PC,Nordborg1976CP,Evans1976R,Behrens1977NSE,Fursov1977AE,Varnagy1982NIM,Androsenko1983CP,Manabe1988FC,Meadows1988ANE,Jingwen1989CJNP,Lisowski1991CP,Shcherbakov2002JNST,Tovesson2014NSE}.  Data is compared to the ENDF/B-VIII.$\beta$5~\cite{ENDFB8B5} evaluation, shown with the evaluated uncertainty.  
The uncertainty at $20$~MeV, the maximum value at which an uncertainty is given, is used for energies greater than that value.
An expanded view of the data is shown in the inset, compared to the ENDF/B-VII.1 \cite{Chadwick2011NDS} and ENDF/B-VIII.$\beta$5 evaluations, indicating a recent 40\% change to the $^{238}$U(n,f) cross section at $1.2$~MeV.}
\end{figure*}

The NIFFTE Collaboration aims to measure the $^{239}$Pu(n,f)/$^{235}$U(n,f) cross section ratio to a total uncertainty of $<$1\% using the fissionTPC.  Previous measurements have reported uncertainties of a similar magnitude~\cite{Staples1998NSE}, but the scatter amongst these suggests that one or more systematic uncertainties may have been unrecognized or underestimated.  The additional information provided by the fissionTPC enables an independent measurement intended to resolve these discrepancies and improve the quality and reliability of the derived nuclear data. Cross section measurements with $^{239}$Pu targets are  more challenging than many other actinides, since the short $^{239}$Pu half-life ($24,110$~years) results in high $\alpha$-particle activity that can lead to significant event pile-up. 

The normalized $^{238}$U(n,f)/$^{235}$U(n,f) cross section ratio presented here has been measured with the fissionTPC in order to demonstrate the measurement technique using this new instrument and quantify sources of systematic uncertainty without the presence of a large $\alpha$-decay background.  
This work presents the energy dependence of the neutron-induced cross section ratio normalized to the ENDF/B-VIII.$\beta$5 evaluation~\cite{ENDFB8B5} at $14.5$~MeV.  Calculation of an absolute normalization was not possible in this work due to the large uncertainties in the neutron beam flux introduced by the chosen target geometry, as described in Section~\ref{sec:Discussion}.

The $^{238}$U(n,f)/$^{235}$U(n,f) cross section ratio is a valuable reference, since the $^{238}$U(n,f) cross section is a standard used in neutron flux measurements \cite{Chadwick2011NDS}. Errors in this ratio can therefore produce correlated errors between different nuclear data sets.  A comparison of past data~\cite{White1967JNE,Stein1968CP,Meadows1972NSE,Meadows1975NSE,Meadows1975PC,Nordborg1976CP,Evans1976R,Behrens1977NSE,Fursov1977AE,Varnagy1982NIM,Androsenko1983CP,Manabe1988FC,Meadows1988ANE,Jingwen1989CJNP,Lisowski1991CP,Shcherbakov2002JNST,Tovesson2014NSE} to the ENDF/B-VIII.$\beta$5 evaluation is displayed in Fig.~\ref{fig:FullU8U5Ratio} along with the evaluated uncertainty.  

A  change was recently made to the $^{238}$U(n,f) cross section evaluation at neutron energies of $\sim1.2$~MeV, as reflected in a comparison of the ENDF/B-VII.1~\cite{Chadwick2011NDS} and ENDF/B-VIII.$\beta$5 evaluations (Fig. \ref{fig:FullU8U5Ratio}, inset).  The 40\% change in the $^{238}$U(n,f) cross section brings the evaluation  closer to recent measurements~\cite{Lisowski1991CP,Shcherbakov2002JNST,Tovesson2014NSE}, but other measurements are scattered between the two evaluations.  The cross section ratio measurement reported here provides  additional input for evaluation of the $^{238}$U(n,f) standard. 

The threshold energy for $^{238}$U(n,f) is $\sim1.2$~MeV, and the $^{238}$U(n,f)/$^{235}$U(n,f) cross section ratio drops dramatically below that energy.  The energy range $0.5 - 30$~MeV was chosen
because many past measurements begin at this same lower bound and the measured ratio uncertainty becomes larger than the ratio at this energy.  The upper bound was chosen because the primary applications of this work do not require measurement at higher energy, the $30$~MeV is the maximum energy reported in ENDF/B-VIII.$\beta$5 for the $^{235}$U(n,f) and $^{238}$U(n,f) cross sections, and wrap-around corrections grow at larger neutron energies.

The following sections of this paper describe the $^{238}$U(n,f)/$^{235}$U(n,f) cross section ratio measurement using the fissionTPC.  Section \ref{sec:Experiment} describes the experimental conditions of the measurement, including the detector, beam properties, and data acquisition.  Section \ref{sec:Data} provides details of the methods used to extract particle information from the recorded data.  In Section \ref{sec:Ratio}, these quantities are combined with a Monte Carlo based efficiency model to generate the cross section ratio, as well as the ratio covariance matrix as a function of neutron energy.

\section{\label{sec:Experiment}Experiment}

The $^{238}$U(n,f)/$^{235}$U(n,f) cross section ratio was determined by measuring fission fragments from half-disk targets of $^{238}$U(n,f) and $^{235}$U(n,f) on a thin 100 $\mu$g/cm$^2$ carbon backing. The detector was operated on the 90L beam line of the Weapons Neutron Research (WNR) facility at the Los Alamos Neutron Science Center (LANSCE)~\cite{Lisowski2006NIMA}, where an 800-MeV proton accelerator provided 125 ps micropulses which are spaced $\sim$1.8 $\mu$s apart.  There were 348 micropulses per macropulse, and 100 macropulses per second delivered to the unmoderated tungsten WNR neutron production target.  The energy of fast neutrons produced via spallation was determined via neutron time-of-flight (nToF).  A steel pipe with a 2 cm inner diameter collimates the neutron beam.

\subsection{\label{sec:FissionTPC}FissionTPC}

The fissionTPC is a two-volume MICROMEGAS TPC, operated using a mixture of argon and isobutane, with an actinide target mounted to the central cathode~\cite{Heffner2014NIMA}.  Ionization electrons produced by charged particle energy loss are drifted away from the cathode by an applied electric field, inducing a current signal in the cathode. 
Collection of ionization charge on a two-dimensional array of readout pads allows $x-y$ reconstruction of interaction positions, while the relative charge arrival time provides the position along the $z$-axis (neutron beam direction). 
To reduce the readout time and lower the event multiplicity, the $5.4$~cm drift length of the device is significantly smaller than that typical for TPCs
used for high-energy physics experiments. 
Having the actinide targets  deposited on a thin carbon backing enabled fission fragments to travel into either volume, allowing measurement of both fragments and increasing the magnitude of the induced cathode signal.

The fissionTPC drift gas composition (high-purity argon and 5\% isobutane) was chosen because it proved to be  resistant to discharges in the MICROMEGAS when operating in a neutron beam.  The operating pressure of $550$~Torr (73.3 kPa) was selected such that spontaneous decay $\alpha$-particles and fission fragment tracks were fully contained within the active area of the detector volume. At this pressure, a local maximum in the drift velocity would be achieved at an applied drift field of about $200$~V/cm. It is typical to operate TPCs close to this maximum to reduce sensitivity to temperature and pressure fluctuations. However, the ionization charge density produced by fission fragments is significantly greater than is generally observed in light-ion TPCs, and  trapping by the large ion space charge was observed to retard the drift of a significant fraction of the ionization electrons.  These trapped electrons resulted in a large charge tail, which complicated tracking and biased the detected track angle.  
By operating at an increased drift field of
$520$~V/cm this effect was significantly reduced, at the cost of slower drift times and greater potential instability in the drift velocity. 

The MICROMEGAS gain stage at the anode includes a thin mesh separated from the pad plane by $75$~$\mu$m.  The 28 kV/cm electric field in this region is significantly higher than in the drift region, resulting in an avalanche that produces a signal gain of 34 at the pad plane.  The pad plane consists of hexagonal pads of $2$~mm pitch.

\subsection{\label{sec:ActinideTarget}Actinide Target}

The target consists of two half-disks of the actinides $^{235}$U and $^{238}$U formed on a thin carbon backing by vacuum deposition~\cite{Loveland2009JRNC}.  
The activity of both long-lived isotopes can be measured using an autoradiograph, i.e. direct in situ counting of their respective  $\alpha$-decay rates (Section~\ref{sec:TargetIsotopics}).
This procedure is complicated by the presence of shorter half-life uranium isotopes, but the relative amounts of these species can be determined by analysis of  $\alpha$-particle track length distributions.
The $^{238}$U deposit includes measurable  $^{235}$U contamination, which is corrected for in the final ratio analysis.

\subsection{\label{sec:DataAcquisition}Data Acquisition}

Each of the $5952$~pads included in the fissionTPC is recorded by a $50$~MHz digitizer~\cite{Heffner2013IEEE}.  These are arranged in  $192$~EtherDAQ cards of $32$~channels each~\cite{Heffner2013IEEE,Heffner2014NIMA}.  When a  digitizer channel exceeds a specified threshold, event recording commences and does not terminate until the signal falls below threshold.  The cathode signal was recorded by a $1$~GHz digitizer.

\section{\label{sec:Data}Data Analysis}

Analysis of the fissionTPC $^{238}$U(n,f)/$^{235}$U(n,f) data set involves many steps.  Track reconstruction is performed on  voxels generated from pad-plane signals to determine quantities such as energy, length, and direction.
The cathode signal is analyzed to determine neutron time-of-flight (nToF) values, which can be converted into neutron energy.  The target isotopics and overall activity must be determined in order to normalize the cross section ratio and correct for all actinide species present.  A beam-target correction must be generated by examining the spatial overlap of the neutron beam with the target actinide density.  Finally, a wrap-around correction is needed to remove contributions from low-energy neutrons. 
The following sections describe how each of these quantities and corrections is determined.

\subsection{\label{sec:TrackReconstruction}Track Reconstruction}

\begin{figure}[h]
\includegraphics[scale=0.7]{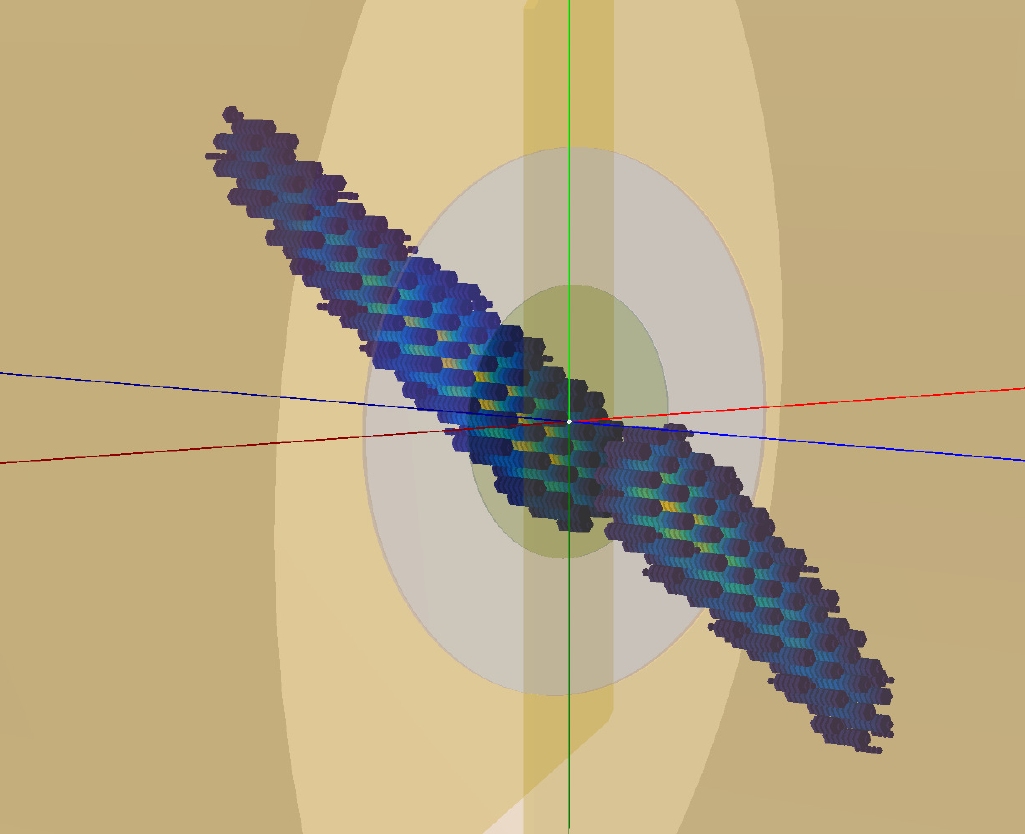}
\caption{\label{fig:FissionVisualization} (Color online) Visualization of fission event in the fissionTPC.  The thin target allows for both fission fragments to be detected, one in each chamber.  The gray disk represents the target holder, and has a 4 cm diameter.  The two fission fragments have a common start vertex, but are displaced in the z-direction to force all voxels of charge into their respective volumes.}
\end{figure}

\begin{figure}[h]
\includegraphics[scale=1.2]{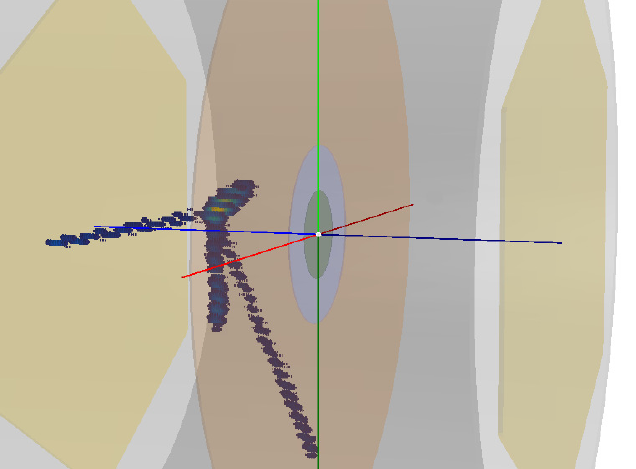}
\caption{\label{fig:CarbonBreakup} (Color online) Visualization of a spallation event in the fissionTPC.  Several light ions can be seen in a single volume all with a common vertex.  The fissionTPC is capable of tracking each particle separately.}
\end{figure}

The first step in the data analysis is to reconstruct the charge clouds recorded by the pad-plane digitizers.  
Since the EtherDAQ front-end performs a charge integration, digitizer waveforms are differentiated using a discrete filter to generate voxels of charge, yielding a three-dimensional representation of the charge cloud detected in the event.
Fission fragments, $\alpha$-particles, recoil protons, and recoil argon and carbon ions can all occur during the same event, even when separated by a significant distance.
For example, Fig.~\ref{fig:FissionVisualization} displays a fission event reconstruction with a  fragment in each volume.  In addition, Fig.~\ref{fig:CarbonBreakup} shows several light ions produced from a spallation event sharing a common vertex.

After the distribution of voxels is generated, tracking algorithms separate individual particles.  The primary tracker used in this work separates non-contiguous charge clouds with an adjacency check.  To increase efficiency, the strips of voxels produced by single pads are combined into columns, before merging with adjacent columns.  The benefits of this tracker are that it is simple, efficient, and  it properly handles most multi-particle events.  After separating the charge clouds,  a track fitter is used to find the track start vertex, end vertex, orientation, and energy.

The track fitting algorithm begins with the assumption that the particle passes through the center of charge of the  cloud, and then finds the axis that minimizes the distance-squared between the axis and each voxel of charge.  The charge is then projected onto that track axis,  the track start and end vertices are found by determining where the charge profile crosses a specific threshold, and then extrapolating back to zero charge.  The threshold is set low enough to primarily be influenced by diffusion.  When identifying the track start and end for fission fragments and $\alpha$-particles, the particles are assumed to travel away from the target plane.  The track fitting threshold depends on diffusion and space-charge effects, and is tuned to the argon/isobutane mixture used for the experiment.  

\begin{figure}[h]
\includegraphics[scale=0.43]{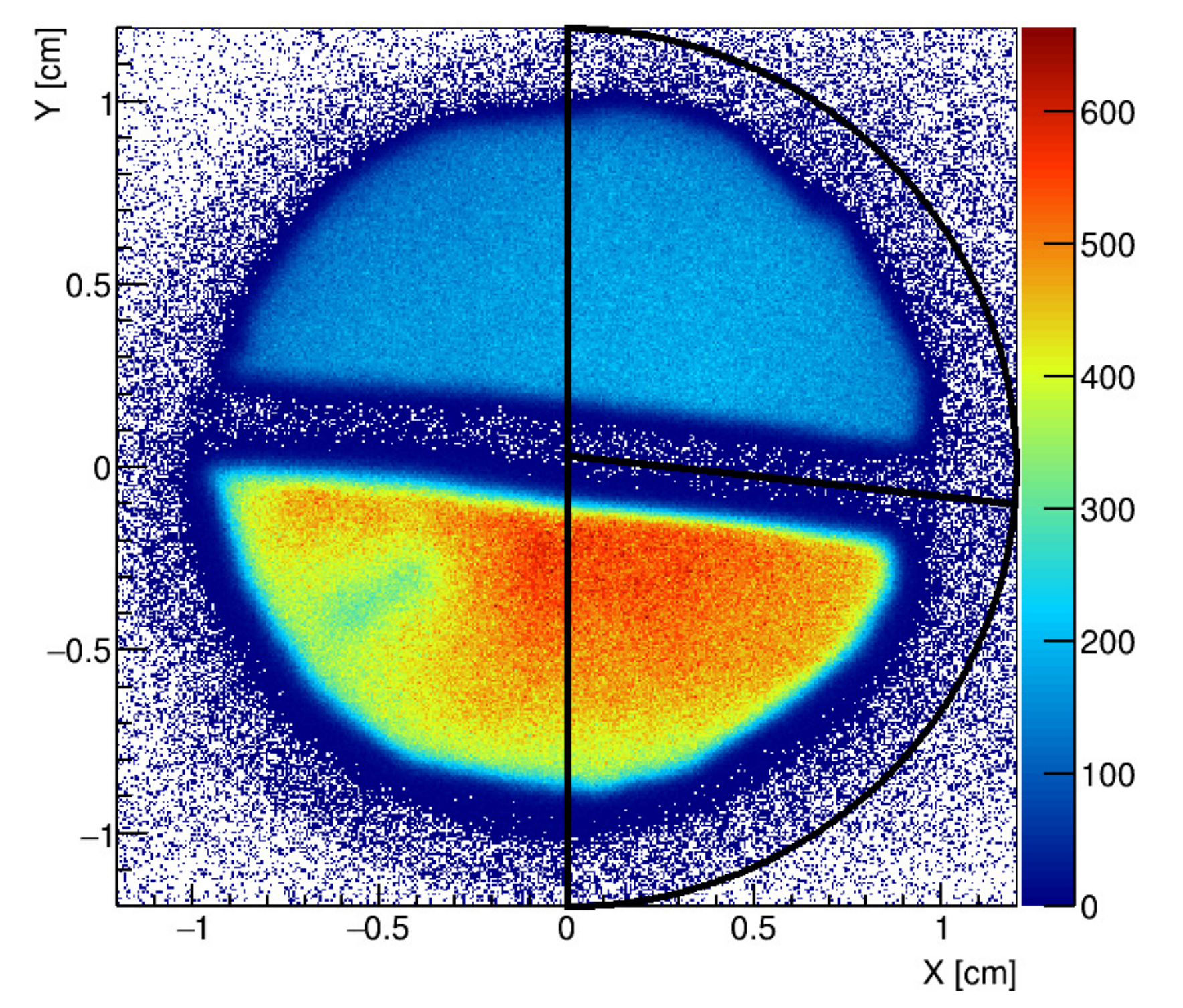}
\caption{\label{fig:TargetCuts} The $x-y$ spatial distribution of fission fragment vertices reconstructed by the fissionTPC. The upper half disk is the $^{238}$U deposit while the lower is the $^{235}$U deposit. Black lines indicate spatial selection cuts: the radial cut prevents backgrounds from actinide contamination on the cathode, while the cut bisecting the two deposits identifies which actinide has fissioned.  Two adjacent missing pixels can be seen near the position (-0.5,-0.4), so only $x>0$ data is considered for the analysis.}
\end{figure}

The $x$-$y$ vertex pointing resolution is $280~\mu$m, as determined from the  spatial distribution  of the actinide deposit edge.  The pointing resolution results in a halo around the target distribution, which can be seen in the fission fragment spatial distribution shown in Fig.~\ref{fig:TargetCuts}.  At larger than 1 cm radius, a background with very low statistics can be seen, which is assumed to result from mis-tracked fragments.  The charge clouds are constructed from data recorded by the anode pixels, and missing pixels bias the spatial profile of the target by lowering some tracks below the fission fragment detection threshold.  Two adjacent missing pixels can be seen in Fig.~\ref{fig:TargetCuts}, and an $x>0$ cut is placed on the data to avoid this bias.

The track fit quality is evaluated by calculating the charge fraction near the fit axis.  Track fits of poor quality typically occur when charge from different particles have spatial overlap. A Hough transform tracking approach is used in such cases~\cite{Hough1959CP,Hough1962Patent}.  The $x-y$, $y-z$, and $x-z$ projections of the three-dimensional charge cloud are analyzed. The line of highest charge density is iteratively removed from the event, with projections being repeated on each iteration.  This has the benefit of cleanly selecting fission fragments, but can result in the splitting of $\alpha$-particles, protons, and recoil ion tracks.  This algorithm is considerably more computationally intensive, and is only used when the initial tracker fails to produce a quality fit.

\begin{figure}[h]
\includegraphics[scale=0.43]{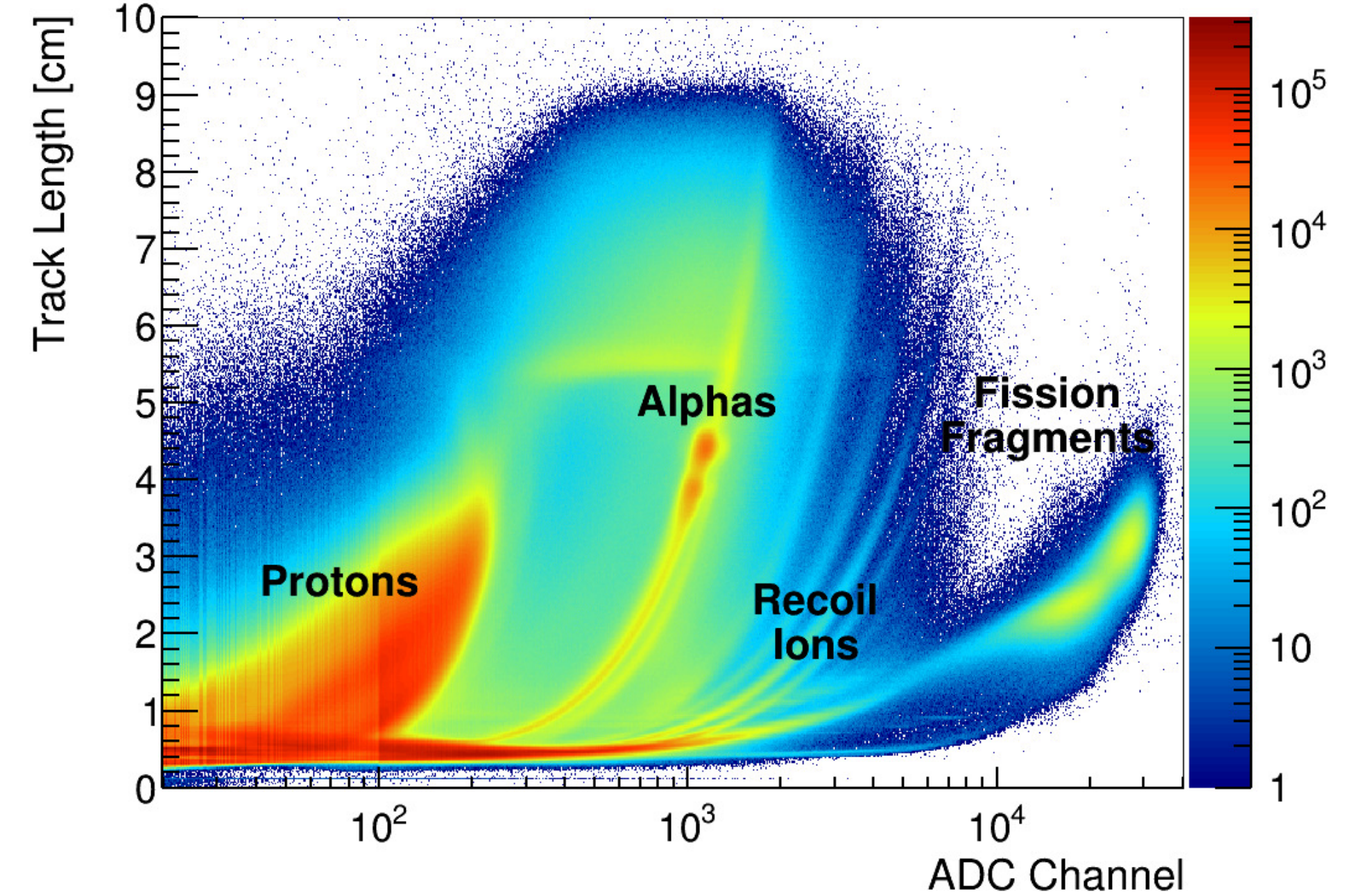}
\caption{\label{fig:EnergyLength} (Color online) Length vs. energy for particles observed in the fissionTPC with the LANSCE neutron beam impinging the device.  ADC channel refers to the uncalibrated particle energy recorded by the digitizers.  Different particles have unique stopping power profiles in the drift gas, and length/energy cuts can be used to isolate specific particle types.  Labels have been added to the different particle distributions. 
}
\end{figure}

Once the fit is complete, the track parameters length and energy can be used to select particles of different atomic mass and atomic number (Fig.~\ref{fig:EnergyLength}).
The large proton flux observed primarily results from $^{1}$H(n,el) in the isobutane, $\alpha$-particles from carbon breakup and $\alpha$-decay, and recoil ions from neutron scattering on carbon and argon.

\subsection{\label{sec:NeutronToF}Neutron Time-of-Flight}

The neutron energy is determined by measuring nToF between the spallation and actinide targets.
An electromagnetic pickup signal provides the start timing reference, while detection of a fission on the fissionTPC cathode provides the stop signal. 
Observation of photofission from $\gamma$-rays produced by spallation of the tungsten target allows for determination of the  propagation delay of the beam between the pickup and the spallation target.
The accelerator micropulses are separated by $\sim$1.8 $\mu$s, and can be combined by accounting for this time difference.

The remaining unknown in the conversion of time to energy is the distance between the tungsten spallation target and actinide target in the fissionTPC.  The nuclide $^{12}$C is known to have large neutron scattering resonances~\cite{Chadwick2011NDS}; insertion of carbon material (a ``carbon filter'') between the production and actinide targets creates notches in the measured nToF distribution at well-known energies. 
The measured nToF, i.e. neutron energy, corresponding to the notch at $2.08$~MeV, in combination with the offset provided by the photo-fission feature, determines the distance between the two targets to be $8.059(3)$~m, where the primary source  of uncertainty was event statistics.

\begin{figure}[h]
\includegraphics[scale=0.43]{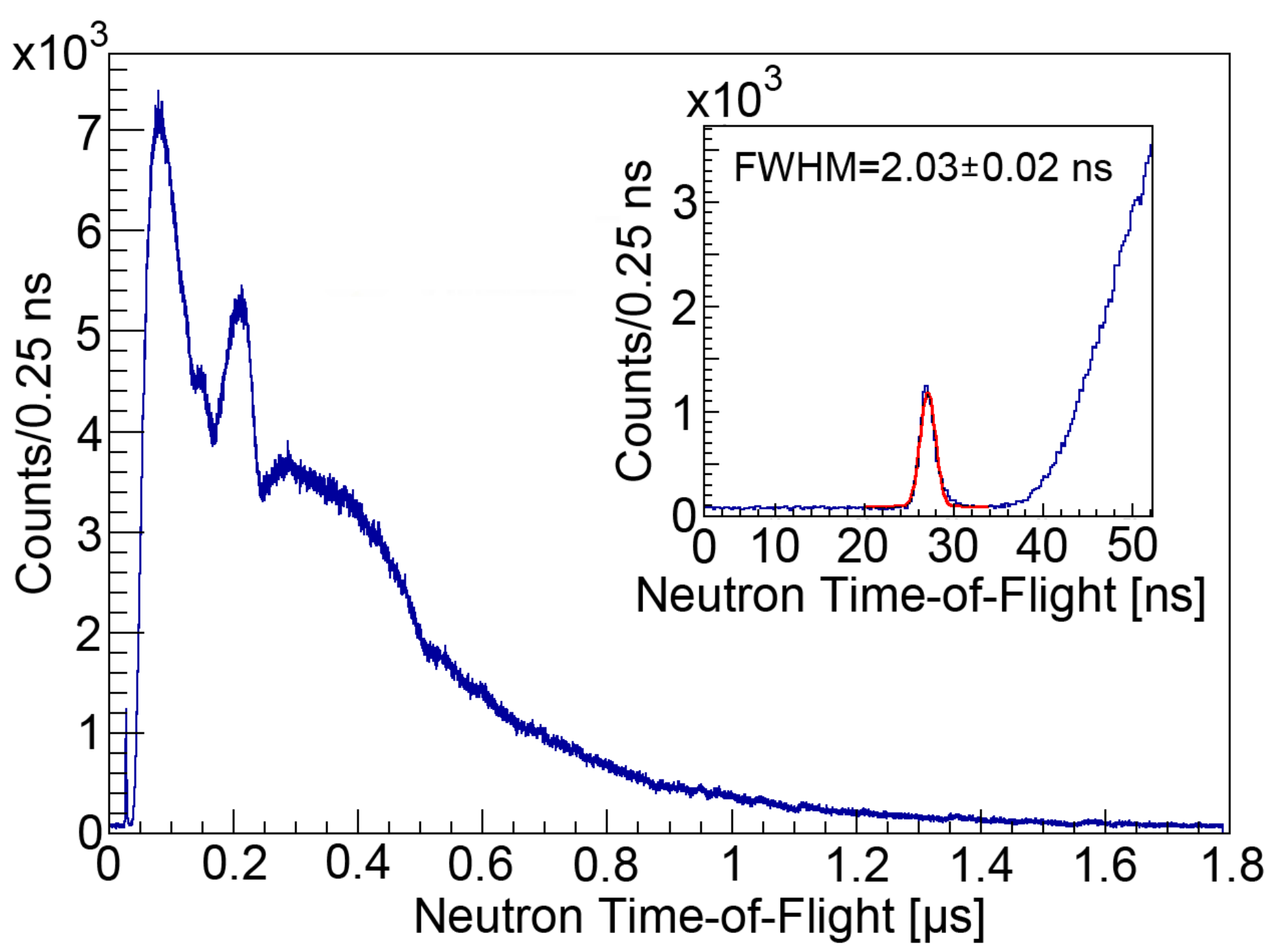}
\caption{\label{fig:nToFShape} (Color online) nToF distribution of the combined $^{235}$U and $^{238}$U targets.  The inset shows a Gaussian fit to the nToF photo-fission distribution, yielding a timing resolution of $2.03(2)$~ns FWHM.
}
\end{figure}

Cathode signal timing is obtained by applying a digital moving-average filter~\cite{Jordanov1994NIMA} and interpolating the rising edge back to the zero-crossing.
The nToF resolution of $2.03(2)$~ns FWHM is  determined by fitting the photo-fission feature with a Gaussian distribution on a flat background (Fig.~\ref{fig:nToFShape}).  
The cathode efficiency relative to the anode for the two actinide deposits was found to be $\sim$99\% for events included in the cross section analysis, with this quantity largely canceling in the cross section ratio.

\subsection{\label{sec:TargetIsotopics}Target Isotopics}

\begin{figure}[h]
\includegraphics[scale=0.43]{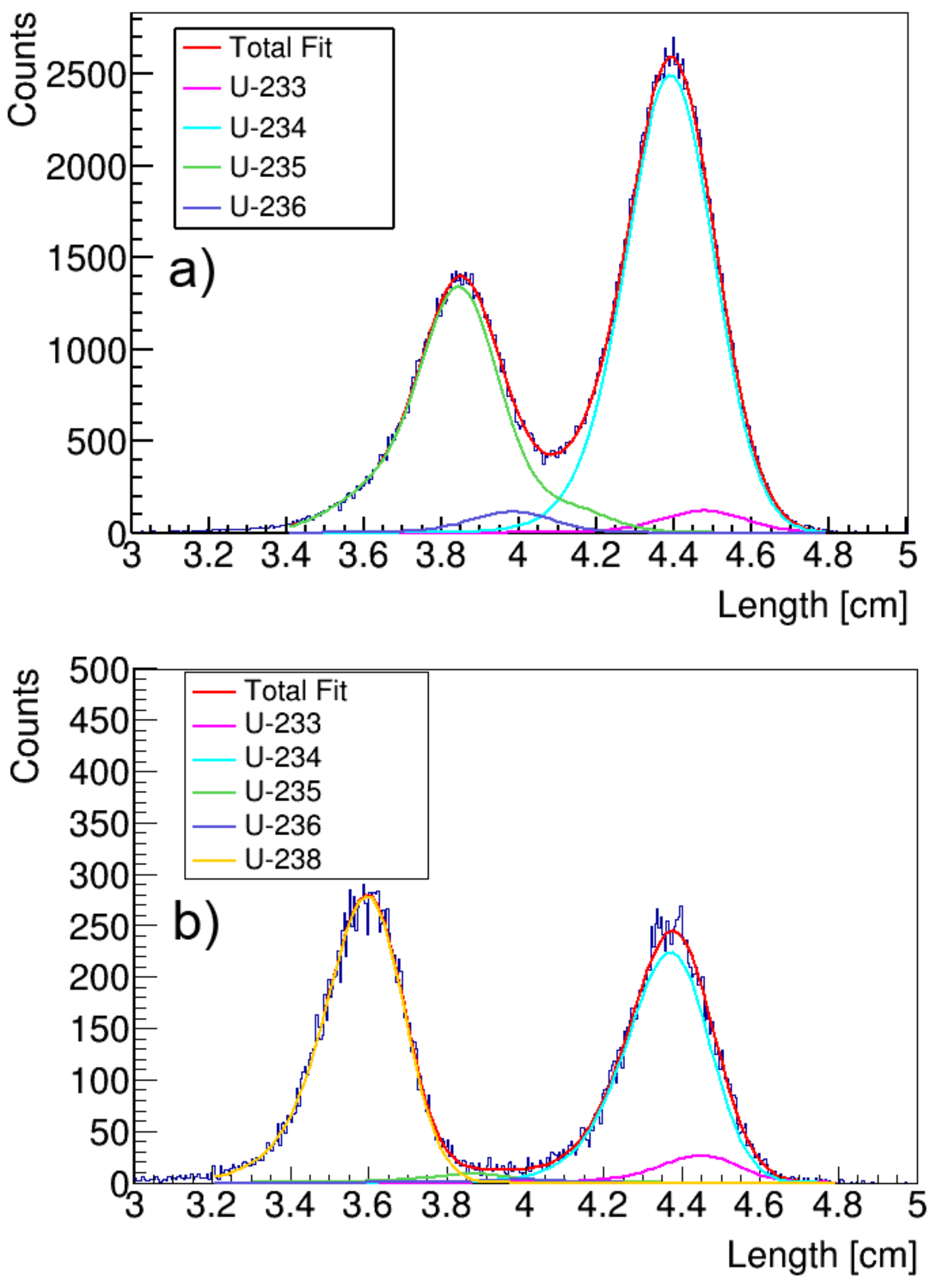}
\caption{\label{fig:UIsotopics} (Color online) Distribution of $\alpha$-particle track lengths emitted by the (a) $^{238}$U and (b) $^{235}$U targets.  Track length is related to  energy by the $\alpha$-particle stopping power.}
\end{figure}

To determine the cross section ratio, it is essential to identify the target atom number of each isotope and correct for any contaminants that could add to the fission fragment count.  
The total $\alpha$-particle activity of each target can be determined by operating the fissionTPC with no incident neutron beam (an autoradiograph).  
The contribution from individual isotopes can be identified through fitting of energy or equivalently, length spectra using the known $\alpha$-particle lines of likely actinide constituents (Fig.~\ref{fig:UIsotopics}).  The length distribution of $\alpha$-particle lines was found to have higher resolution than the particle energy distribution.  
The width and energy scales each have linear calibrations with two free parameters.  A skew term is added to describe energy straggling in the target, resulting in a peak shape that is the convolution of an exponential with a Gaussian.  The peak areas for each isotope are additional free parameters.

Resulting isotopic abundances are given in Table \ref{tab:isotopics}.  
The $^{238}$U target contains $0.57(10)$\%~$^{235}$U, which must be corrected for when calculating the fission cross section.  $^{238}$U has a neutron-induced fission threshold of $\sim$1.2 MeV, and the $^{235}$U contaminant would appear in the cross section ratio as a flat, non-zero value below threshold, due to the contaminant being in ratio with itself.   The $^{235}$U target contains 0.25(4)\%~$^{236}$U, an amount that results in a small fission rate and is not corrected for here.  The ratio of $^{235}$U atoms to $^{238}$U atoms in the respective targets was found to be $0.917(13)$. 

The $^{233}$U and $^{234}$U contaminants in both targets have a negligible effect on the fission cross section ratio due to their small atomic fractions.  The $\alpha$-decay activity from these isotopes is significant due to their short half-lives, and must be accounted for when determining the actinide density of the isotopes of interest: in the $^{235}$U target, the $^{235}$U was found to contribute $35\%$ of the total $\alpha$-decay activity, compared to $50\%$ $^{238}$U $\alpha$-decay activity for the $^{238}$U target.

\begin{table}
\caption{\label{tab:isotopics}Measured isotopic abundances in the two targets.}
\begin{ruledtabular}
\begin{tabular}{lcc}
Isotope & $^{238}$U Target (\%)& $^{235}$U Target (\%) \\
\hline
$^{233}$U & 0.0003(2) & 0.002(1) \\
$^{234}$U & 0.0046(4) & 0.060(2) \\
$^{235}$U & 0.57(10) & 99.69(4) \\
$^{236}$U & 0.005(3) & 0.25(4) \\
$^{238}$U & 99.4(1) &  \\
\end{tabular}
\end{ruledtabular}
\end{table}

\subsection{\label{sec:TargetBeam}Target-Beam Correction}

\begin{figure}[h]
\includegraphics[scale=0.43]{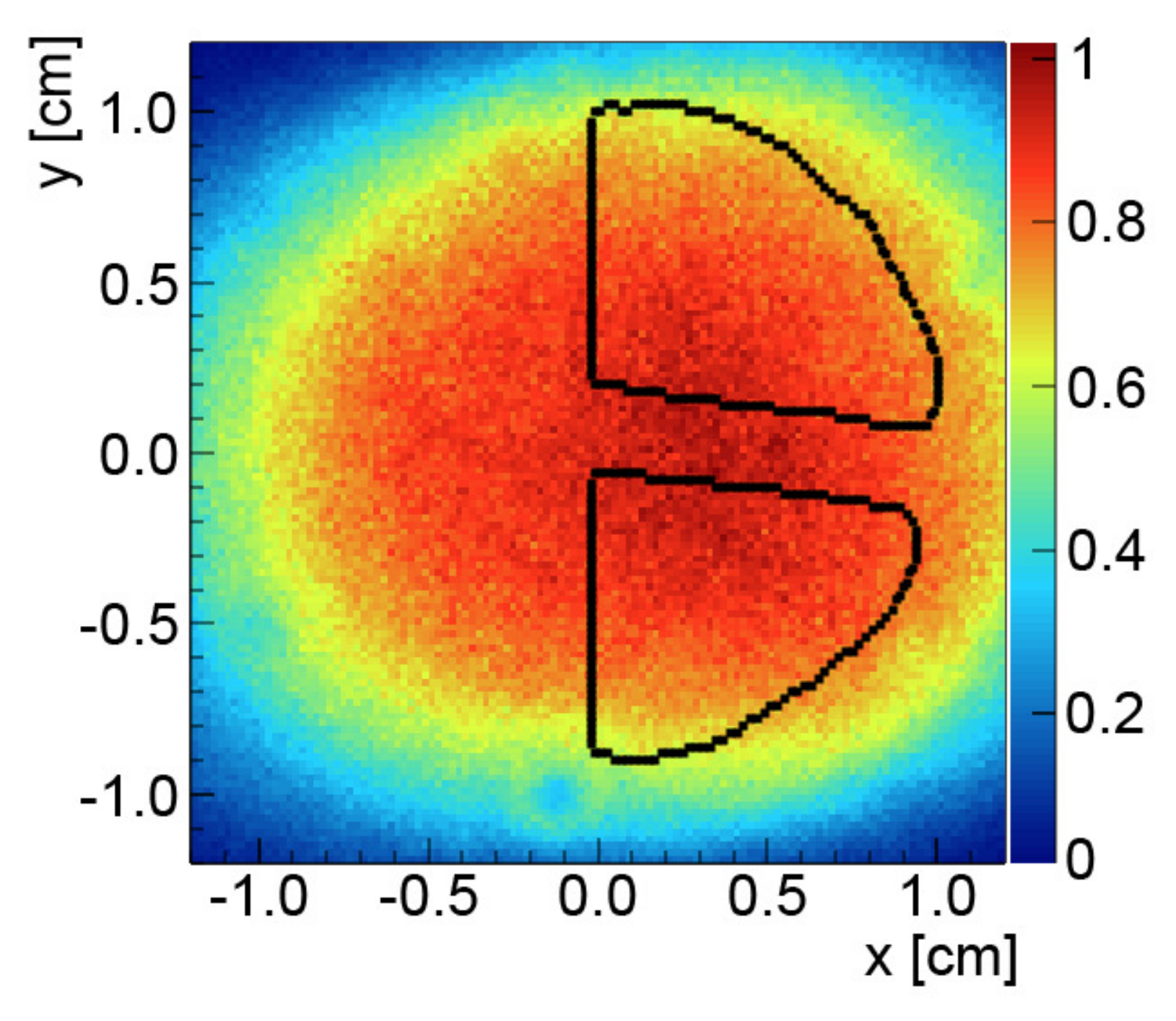}
\caption{\label{fig:BeamFlux} (Color online) Normalized distribution of recoil proton start vertices recorded during high-gain fissionTPC operation, representing the spatial distribution of neutrons incident on the actinide target.  The distribution has been convolved with a Gaussian of $\sigma=0.3$~mm  to reduce aliasing effects from ADC thresholds.  
The black curves outline the regions of actinide target deposit used for the cross section ratio determination, where only $x>0$ is considered due to inactive pad plane pixels.  The aliased shape of the outline represents the binning of the target deposit histogram.}
\end{figure}

\begin{figure}[h]
\includegraphics[scale=0.43]{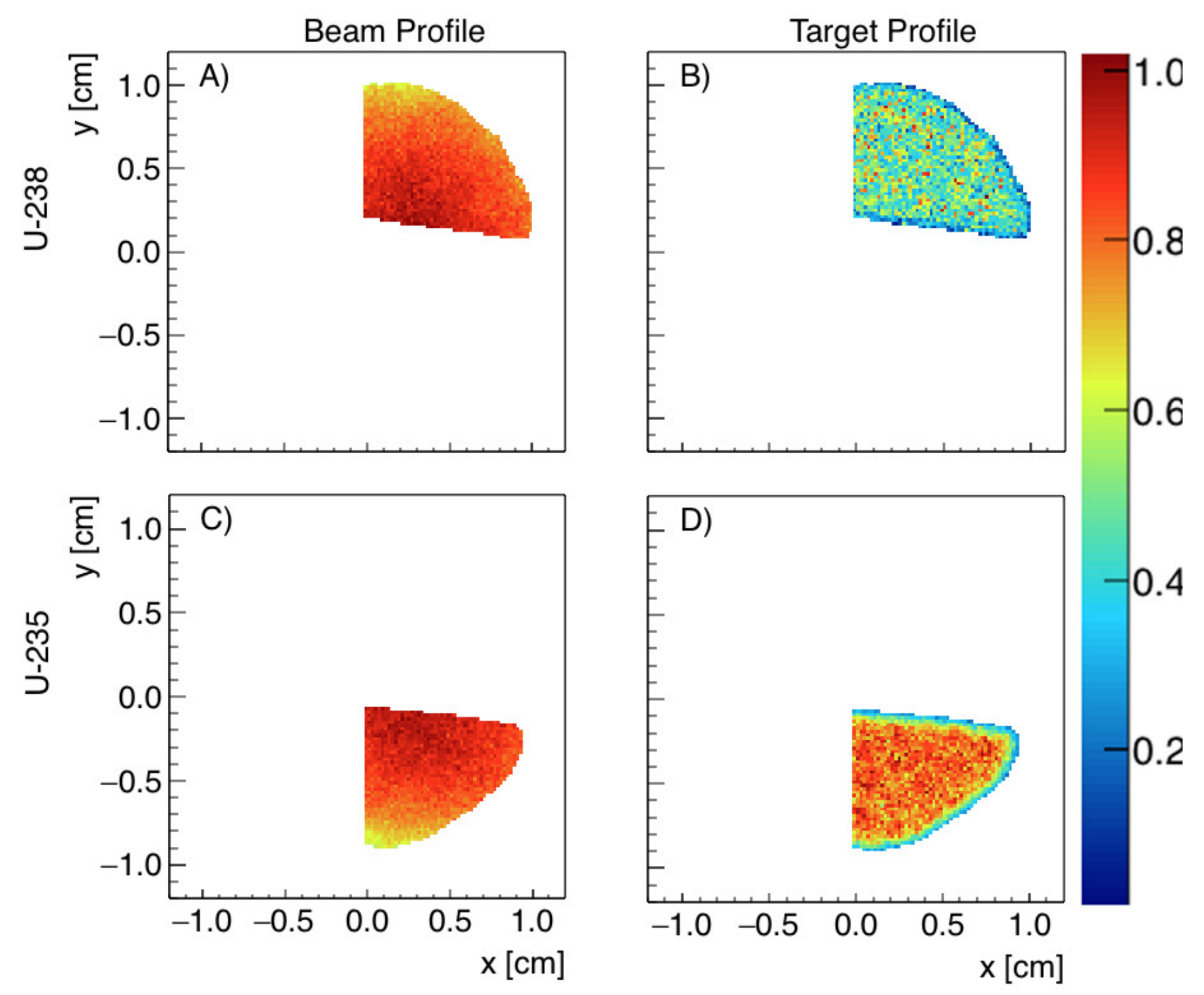}
\caption{\label{fig:BeamTarget} (Color online) Normalized distributions of the beam flux and target activity for the $^{238}$U ((a) \& (b)) and $^{235}$U ((c) \& (d)) actinide deposits.  
Bin-by-bin multiplication of these distributions determines the beam and target overlap.  Only $x>0$ is considered due to inactive pad plane pixels.
}
\end{figure}

The measured fission rate from each target is proportional to the overlap between the spatial distribution of the actinide deposit and the neutron flux.  Recording the start vertex of protons produced by elastic scattering of neutrons on hydrogen in the isobutane gives a measure of the spatial profile of the neutron flux in the fissionTPC (Fig.~\ref{fig:BeamFlux}). In order to record full proton tracks, including the start vertex where the ionization density is lowest, it was necessary to periodically record a subset of data with increased MICROMEGAS gain. The neutron flux spatial profile was found to be static in time, which allows this subset of proton data to be applied to the longer fission measurement.  The beam and target overlap is determined from the product of the normalized neutron distribution and the normalized actinide distribution (Fig. \ref{fig:BeamTarget}).  The ability to measure the spatial dependence of the target and beam overlap in situ is a unique feature of the fissionTPC,  providing the opportunity to study associated systematic uncertainties.

\subsection{\label{sec:WrapAround}Wrap-Around Correction}

\begin{figure}[h]
\includegraphics[scale=0.8]{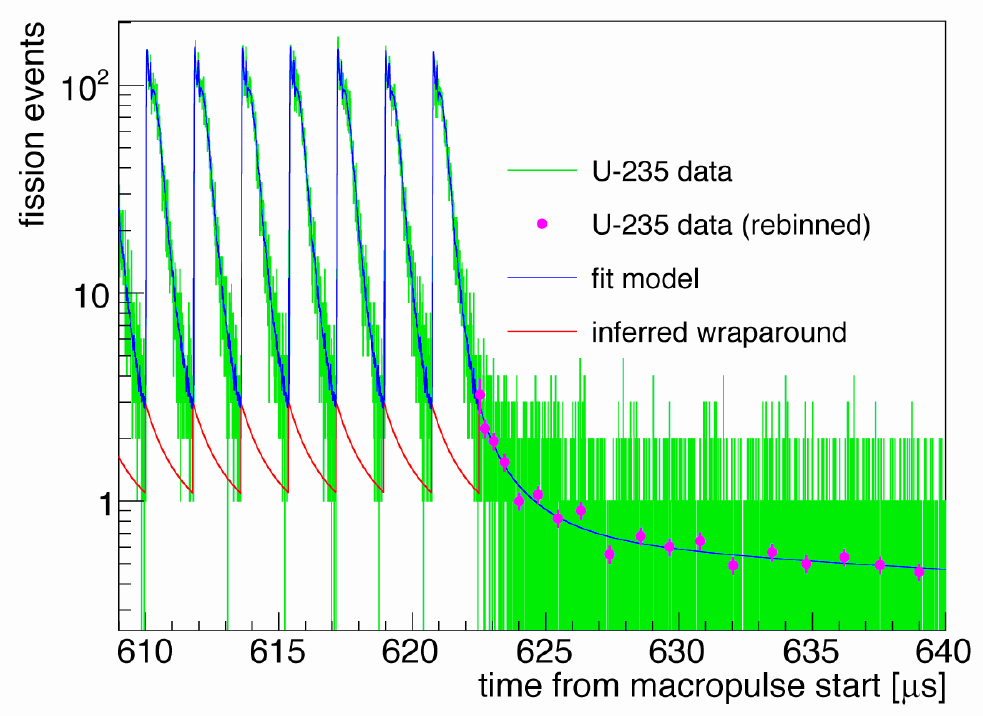}
\caption{\label{fig:WrapExpand} (Color online) Determination of  the wrap-around correction in the $^{235}$U data . The nToF data (green), averaged in the low-energy tail region (magenta), are fit to determine the wrap-around contribution (red line) to the nToF model (blue line).  The nToF model consists of a logarithmic spline following the distribution of prompt neutron data.} 
\end{figure}

\begin{figure}[h]
\includegraphics[scale=0.43]{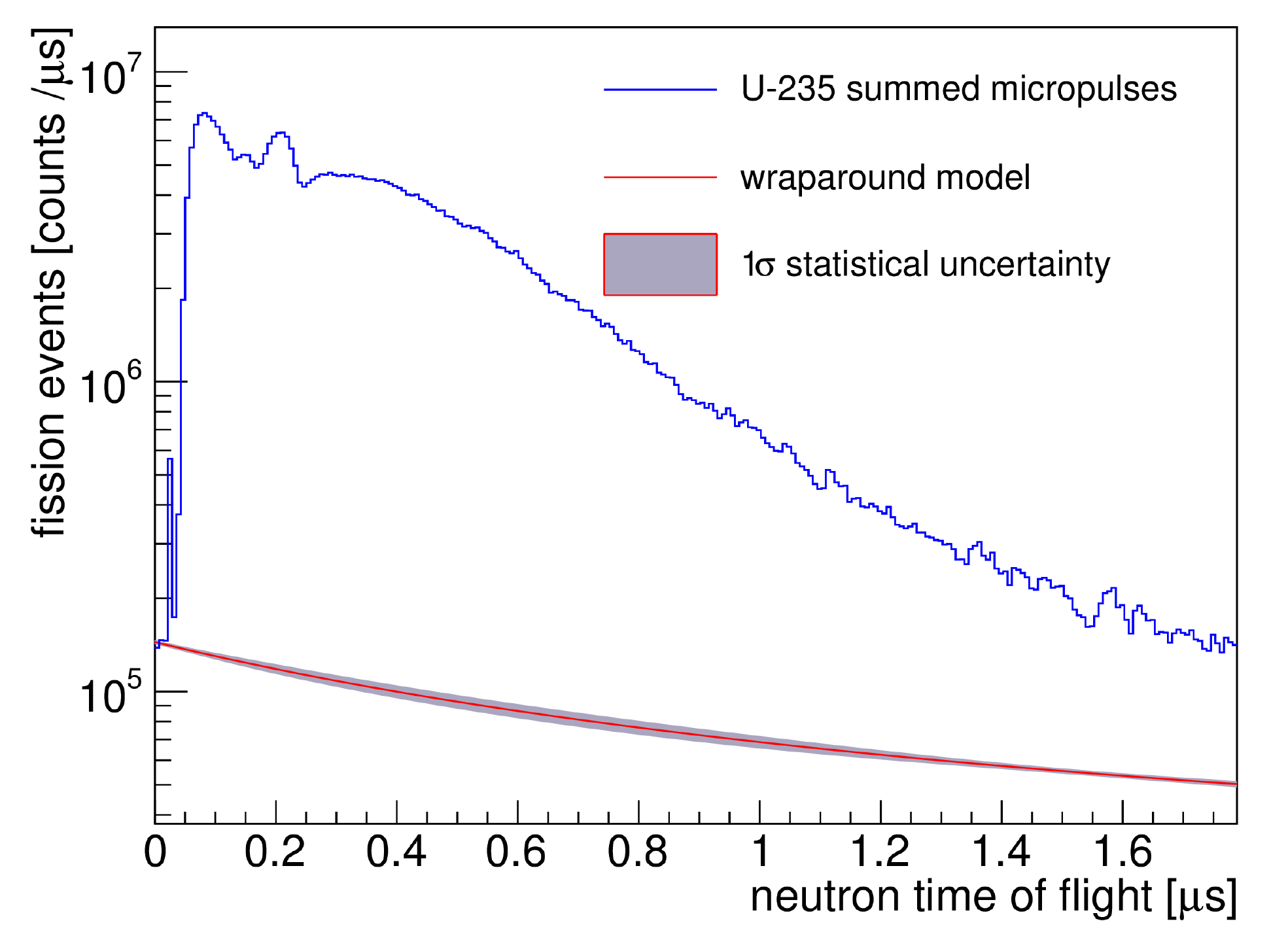}
\caption{\label{fig:WrapCombined} (Color online) The nToF data and wrap-around correction after all micropulses are combined.  The band around the red line represents the uncertainty of the wraparound fit, which was produced by propagating the fit covariances.}
\end{figure}

The LANSCE proton accelerator produces bunches spaced $\sim$1.8~$\mu$s apart, and low-energy neutrons from one bunch may carry-over to later bunches.  This results in low-energy contributions to the high-energy region of the time-of-flight distribution, which must be subtracted.  The nToF distribution represents the product of the neutron flux with the fission cross section, converted from energy to time.  
Without nToF measurements taken with a larger bunch spacing, this distribution can be difficult to determine.

The recorded data continues $\sim 70~\mu$s beyond the last micropulse, allowing the wrap-around contribution to be determined via a fitting procedure that must also account for  contributions from previous micropulses.  
A logarithmic spline was used to describe the wrap-around contribution, as illustrated for $^{235}$U data in Fig.~\ref{fig:WrapExpand},
since this was found to describe the low-energy tail of the nToF distribution well over large time scales.  
Combining the micropulses produces a total nToF distribution (Fig.~\ref{fig:WrapCombined}).  The fit parameter covariance matrix is used to generate Monte Carlo variations of the fit parameters, which are interpreted as the uncertainty band associated with the wrap-around fitting procedure.

\section{\label{sec:Ratio}Cross Section Ratio}

The cross section ratio measured here is defined by Eq. \ref{eq:CSRatio}, where $x$ refers to the unknown and $s$ refers to the standard actinide.  In this case, $^{238}$U is considered the unknown and $^{235}$U the standard:

\begin{multline}
\frac{\sigma_x}{\sigma_s}=\frac{\epsilon^s_{ff}}{\epsilon^x_{ff}}\cdot\frac{\Phi_s}{\Phi_x}\cdot\frac{N_s}{N_x}\cdot\frac{\Sigma_{XY}(\phi_{s,XY}\cdot n_{s,XY})}{\Sigma_{XY}(\phi_{x,XY}\cdot n_{x,XY})}\cdot\frac{w_s}{w_x}\\
\cdot\left(\frac{(C^x_{ff}-C^x_{r}-C^x_{\alpha})-C^x_{bb}}{(C^s_{ff}-C^s_{r}-C^s_{\alpha})-C^s_{bb}}-G^{sx}_{ss}\right)
\label{eq:CSRatio}
\end{multline}

In this equation, $\epsilon_{ff}$ refers to fission fragment detection efficiency, which will be described in Section \ref{sec:Efficiency}.  
${\Phi_s}/{\Phi_x}$ represents the neutron flux ratio, which was found to be $1.028(1)$ using the proton spatial profiles shown in Fig.~\ref{fig:TargetCuts}. 
${N_s}/{N_x}$ is the number ratio of the two actinides, which was shown to be $0.917(13)$ in Section \ref{sec:TargetIsotopics}. 
${\Sigma_{XY}(\phi_{s,XY}\cdot n_{s,XY})}/{\Sigma_{XY}(\phi_{x,XY}\cdot n_{x,XY})}$ is the beam and target overlap term, which was found to be $1.002(7)$.
${w_s}/{w_x}$ refers to the detector live time ratio, which is estimated to be $100\%$.

The  $C$ terms refer to events detected per neutron energy bin. $C_{ff}$ is the number of fission fragment counts in an energy bin after particle identification cuts are applied.  $C_r$ is the estimated number of background recoil events misidentified as fission fragments.  $C_\alpha$ is the estimated number of pile-up $\alpha$-particle events misidentified as fission fragments.  $C_{bb}$ is the wrap-around correction factor, which is fit for both $^{235}$U and $^{238}$U.  $G^{sx}_{ss}$ refers to the ratio of the number of atoms of isotope $s$ found in deposit $x$ to the number of atoms found in deposit $s$.  This is a contaminant correction for the presence of $^{235}$U in the $^{238}$U target, and is found to be $0.63(10)\%$.

Although Eq.~\ref{eq:CSRatio} is formulated to produce an absolute cross section ratio, the ratio reported here is normalized to the ENDF/B-VIII.$\beta$5 evaluation at 14.5 MeV. 
An absolute normalization was not possible for this measurement due to a large normalization uncertainty ($\sim$10\%) resulting from the separated actinide deposits.  Difficulties associated with mapping the proton beam flux shown in Fig.~\ref{fig:TargetCuts} to a neutron flux are assumed to be the cause.  
Future work will use thick backed targets with back-to-back actinide deposits that will allow multiple neutron beam flux measurement methods.  With a back-to-back target, any spatial flux variations are common to both targets.  
Additionally, the neutron beam spatial distribution can be measured directly by dividing the fission distribution by the actinide density.

\subsection{\label{sec:FissionCuts}Fission Fragment Selection Cuts}

\begin{figure}[h]
\includegraphics[scale=0.43]{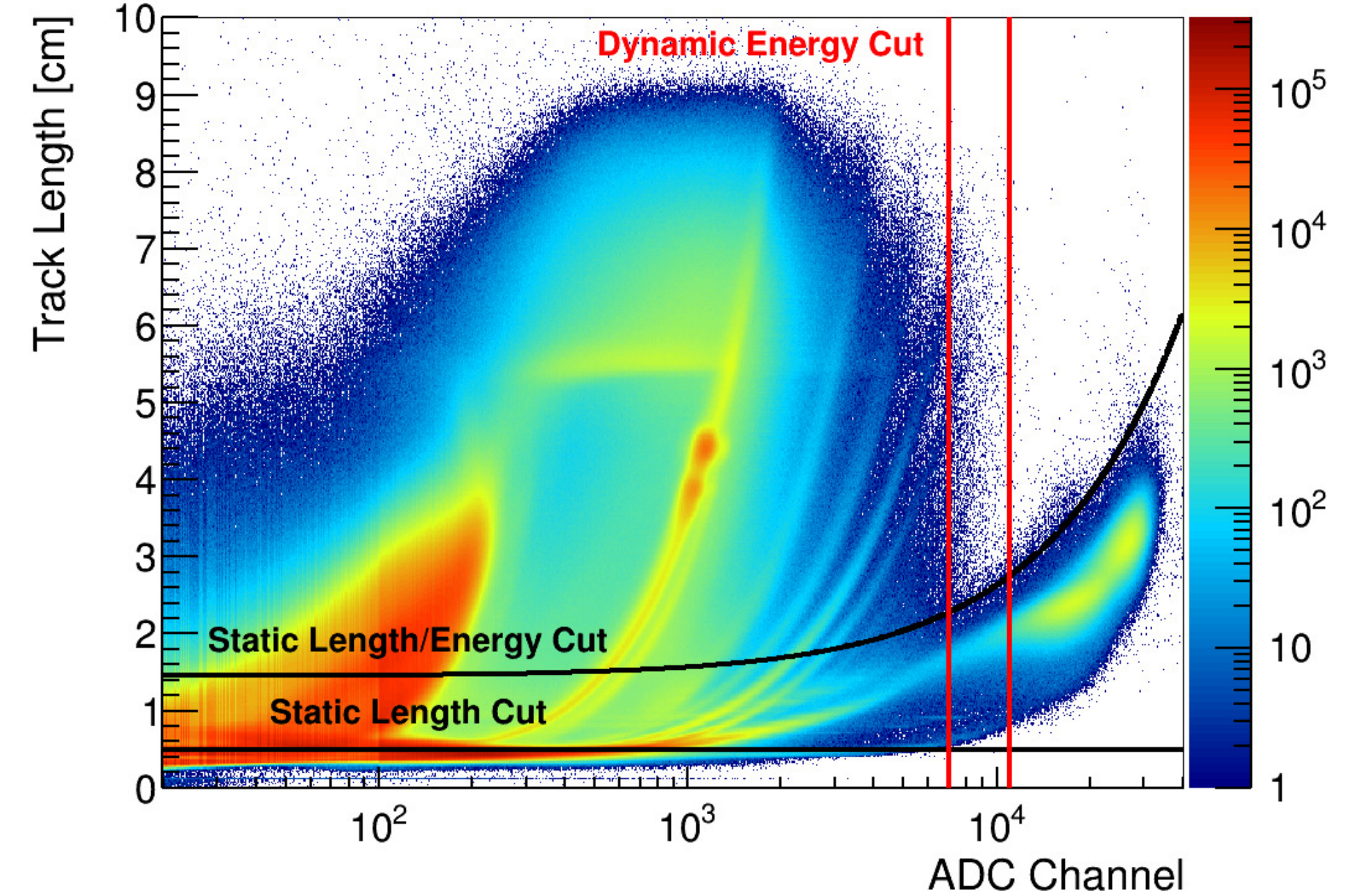}
\caption{\label{fig:PIDCuts} Selection cuts are applied to the energy vs. length distribution of detected particles.  There are two static cuts which remove background particles, and one dynamic cut (varied within the range shown by vertical red lines) which is used to determine residual uncertainties from the efficiency correction.  ADC channel refers to the uncalibrated particle energy recorded by the digitizers.  
}
\end{figure}

The fission fragment detection efficiency  term and all of the $C$ terms in Eq. \ref{eq:CSRatio} depend strongly on the particle identification (PID) cuts that are applied to the data set. 
By definition, as different event selection criteria are applied to the fissionTPC data, $C_{ff}$ and $\epsilon_{ff}$ will change proportionately if the efficiency term is calculated correctly.  
Visual representations of the fission fragment selection cuts applied are displayed in Fig.~\ref{fig:TargetCuts} (spatial actinide selection) and Fig.~\ref{fig:PIDCuts} (particle identification).
Two particle identification static cuts remove non-fragment background, while a dynamic cut is used to estimate residual uncertainties in the fission fragment detection efficiency ($\epsilon_{ff}$) determination process.  The dynamic cut is sampled over a range of different values.
Since $\epsilon_{ff}$ represents the fraction of fission fragments that are observed in the detector with all selection cuts applied, any observed variation in the cross section ratio as the dynamic cut varies is considered to be a residual uncertainty in the determination of $\epsilon_{ff}$ itself.

\begin{figure*}[t!]
\includegraphics[scale=0.43]{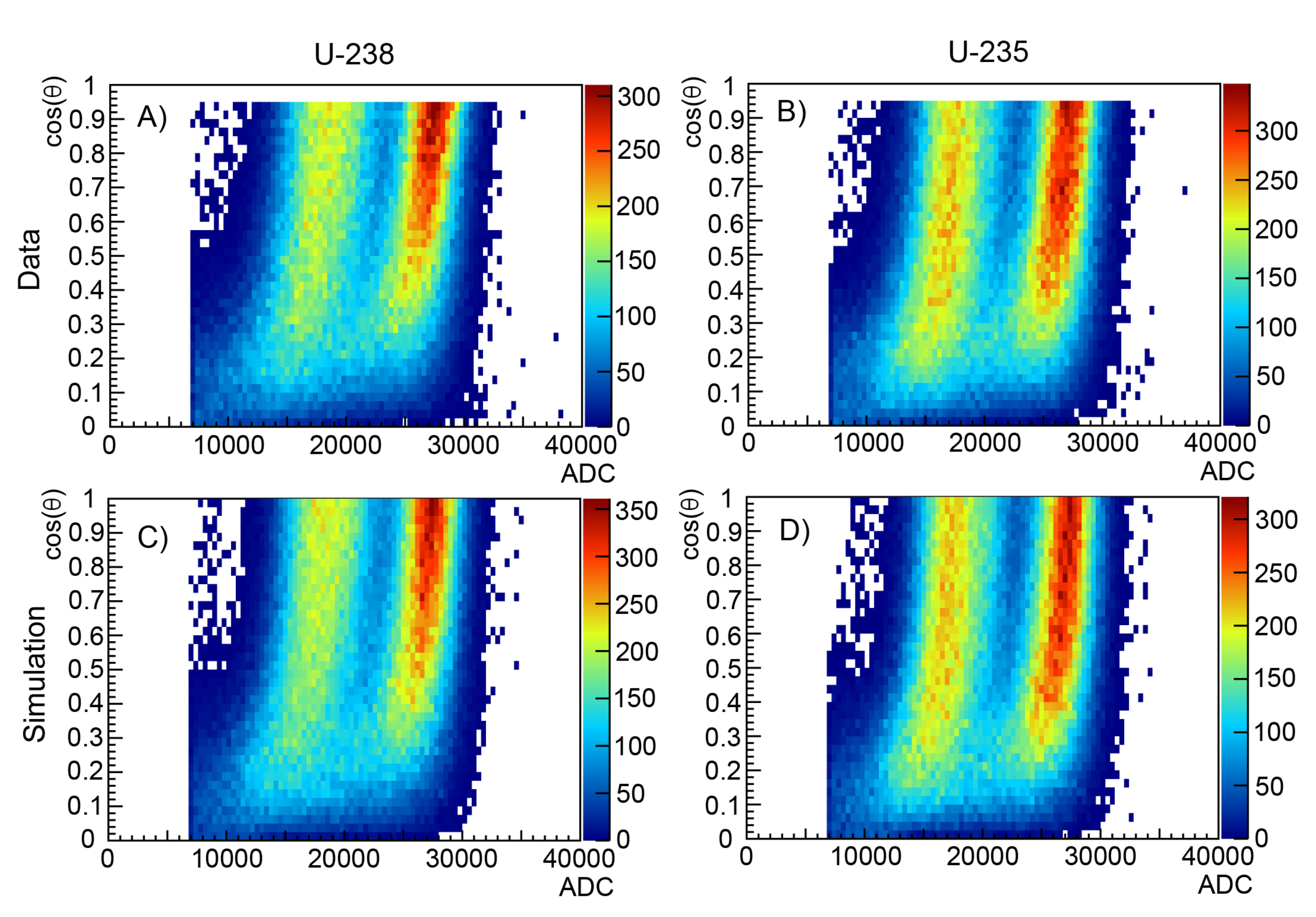}
\caption{\label{fig:EfficiencySimCompare} Energy of observed fission fragments as a function of emission angle from the target ($\cos(\theta)$, $\cos(\theta)=1$ is emission perpendicular to the target). ADC refers to the uncalibrated particle energy recorded by the digitizers.  These distributions are compared for data (A \& B) and simulation (C \& D) for both isotopes.
The two vertical bands represent the escape angle distributions of light and heavy fission fragments.  
These distributions are used to determine $\epsilon_{ff}$ via a fitting procedure with a Monte Carlo simulation.
The data excludes $\cos(\theta)>0.95$ to avoid electronics saturation effects.
Different neutron energy ranges are displayed for the two actinides:  $^{238}$U between $1.33-2.51$~MeV and $^{235}$U between $0.16-0.42$~MeV.
}
\end{figure*}

The dynamic energy cut varies over the range shown in Fig. \ref{fig:PIDCuts}: the low-energy limit is chosen to eliminate the vast majority of non-fragment background, while the high-energy limit removes a small fraction of low-energy fission fragments.
Through careful selection of these limits, the $C_r$ and $C_a$ terms of Eq. \ref{eq:CSRatio} are rendered negligible.  
However, increasing the energy threshold for fission fragment detection has the consequence of increasing the relative uncertainty of $\epsilon_{ff}$.

\subsection{\label{sec:Efficiency}Fission Fragment Detection Efficiency}

The efficiency with which the fissionTPC experimental configuration detects fission fragments, $\epsilon_{ff}$, is clearly of central importance to any fission cross section ratio measurement. The detailed event-by-event information captured by the fissionTPC is used to build and tune a complex phenomenological
efficiency model as a function of incident neutron energy. The efficiency model captures a myriad of transport and loss effects, in addition to underlying nuclear data and the analysis selections described in Section~\ref{sec:FissionCuts}. 
Processes and parameters that have empirically been found necessary to represent the fissionTPC data include fission product yields, fission fragment stopping power, quantum and kinematic anisotropy, and target thickness, composition, and surface roughness. 
Monte Carlo simulations of these effects are used to implement the efficiency model, with the required parameters being determined by fitting observable distributions to fissionTPC data. 
This method is computationally intensive ($\sim 2000$~CPU hours for the final efficiency calculation) since a Monte Carlo realization must be generated for each parameter set. However, there is no analytical approach of which we are aware for this complex problem. 

Changes in the fission fragment detection efficiency, $\epsilon_{ff}$, as fragment energy selection cuts are applied is primarily caused by variable energy loss in the target as a function of emission angle ($\cos(\theta)$, $\cos(\theta)=1$ is emission perpendicular to the target).  
When a fission fragment escapes from the target traveling perpendicular to the target plane, there is minimal energy loss.  
When a fission fragment travels parallel to the target plane, significantly more energy loss can occur, which can result in the fragment stopping in the target and being undetected.  
Furthermore, the minimum energy selection cut displayed Fig.~\ref{fig:PIDCuts}  can result in additional fission fragment losses.  
These losses can be observed by examining the relationship between emission angle and energy (Fig.~\ref{fig:EfficiencySimCompare}), where the fission fragment distributions trend towards lower energy at smaller values of $\cos(\theta)$. 
This emission angle versus fragment energy distribution is the primary representation of fissionTPC data that we use to build and constrain the efficiency model. As we will describe, features in this distribution are sensitive to a number of experiment parameters that are otherwise difficult or impossible to access. 

To better highlight these features, a neutron energy selection has been applied to the distributions shown in Fig.~\ref{fig:EfficiencySimCompare}: for $^{238}$U neutron energies between $1.33-2.51$~MeV are displayed while for $^{235}$U the range is $0.16-0.42$~MeV .
At higher energies, fission anisotropy and the kinematic boost from the incident neutron energy cause forward peaking in the fission fragment angular distribution, so the energy selection upper bound is kept as low as possible while maintaining adequate statistics for the efficiency modeling procedure.  
Because of the  $^{238}$U fission threshold, this target is sampled at higher energy than that for $^{235}$U.
Fission fragment angular distributions in the fissionTPC have been previously studied in detail over a range of incident neutron energies~\cite{Kleinrath2016Thesis}.  
At very forward angles ($\cos(\theta)>0.95$) saturation of pad-plane amplifiers occurs since such tracks occupy few pad-plane pixels. Accordingly, such tracks are excluded from the efficiency modeling procedure.

We use Monte Carlo simulation to recreate the measured $\cos(\theta)$ vs. energy distribution. 
The parameters required are found by performing a multi-dimensional fit to minimize a $\chi^2$ comparison of data and the Monte Carlo representation. 
We build the Monte Carlo simulation by considering fission fragment transport from the target into the active region of the TPC.  The Fission Product Yields (FPY) for each neutron energy bin are determined using the energy of forward-escaping particles in the data ($\cos(\theta)$ between $0.775 - 0.975$), which have minimal energy loss.  The approximate fragment mass is calculated kinematically using the fragment energy and total actinide mass.  The small amount of energy straggling for forward-traveling fragments is corrected for in the FPY determination by deconvolving the estimated energy loss in the target.  During fission fragment transport, energy loss of these particles traveling through the target is determined using parametrized stopping power functions derived from SRIM~\cite{Ziegler2010NIMB}.
The validity of SRIM stopping powers for fission fragments in thin foils was previously studied, and roughly mass-independent differences of up to 30\% were found \cite{Knyazheva2006}.  For the efficiency model in this work, these differences are correlated with the target thickness fit parameter, and should not impact the calculated efficiency.

\begin{figure}[h]
\includegraphics[scale=0.38]{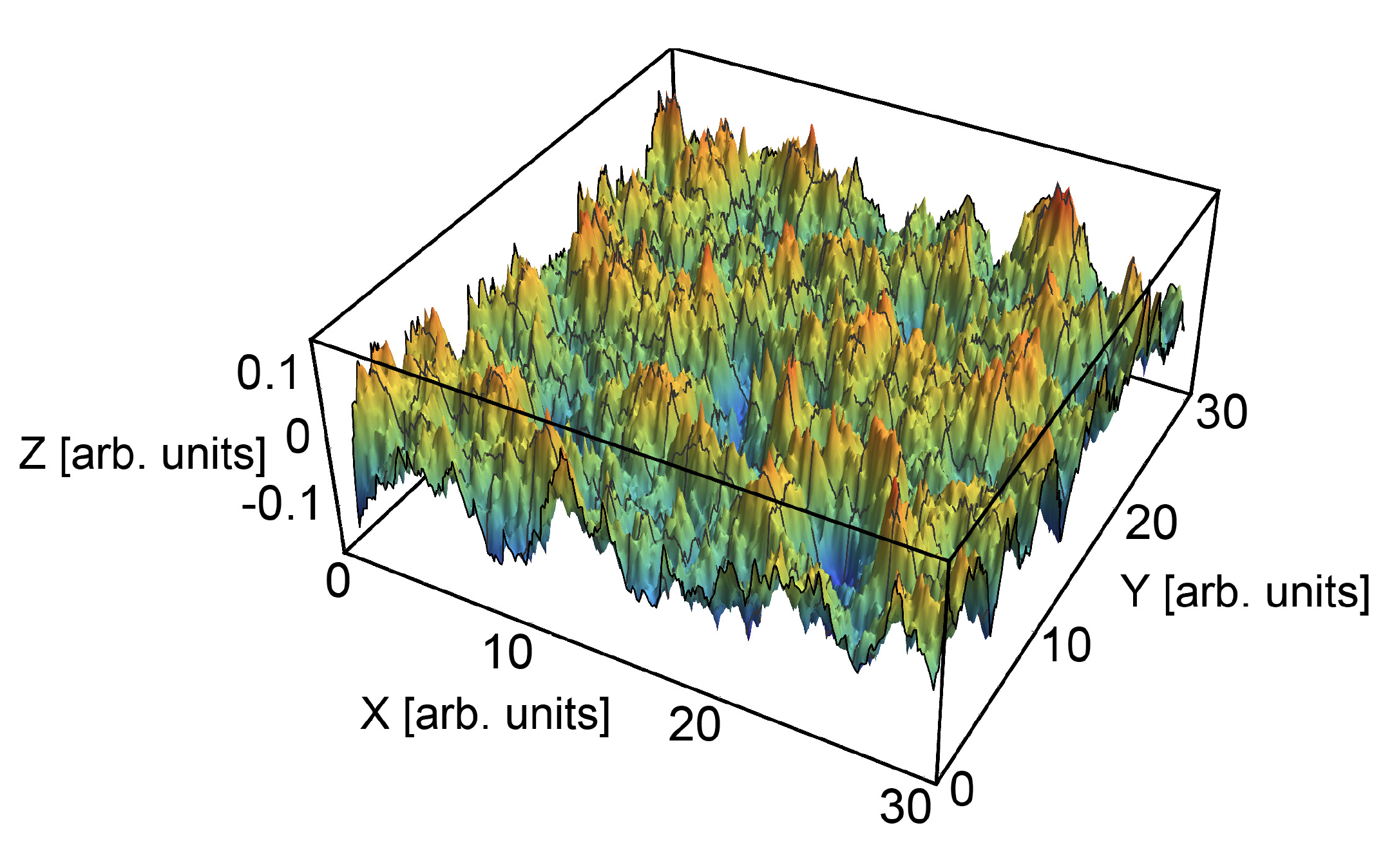}
\caption{\label{fig:SimRoughness} Representative roughness distribution of the actinide targets calculated with a fractal noise model.  The axis units are arbitrary, but are common for all axes.  The ratio of the height to the wavelength determines the surface normal distribution.}
\end{figure}

Target roughness must also be considered to account for the difference between the surface normal and the TPC drift field direction.  Past work with molecular-plated targets on thick backing \cite{Sadi2011NIM} revealed short-wavelength roughness ($\sim$5 $\mu$m), but in the case of a thin carbon backing longer wavelengths are expected \cite{Henderson2011NIM}.
The surface roughness for this work is represented by a simple fractal noise model, generated by combining Perlin noise fields \cite{Perlin1985SCG}, and Fig. \ref{fig:SimRoughness} displays a representative target roughness distribution.  The axis units are arbitrary, but are common for all axes.  
Sampling the surface normal distribution yields a $\cos(\theta)$ distribution with the form $\exp(x/\beta-1)$ where the parameter $\beta$ represents the roughness.  
Having found this simple representation of the target roughness, we have similarly investigated the effect of that roughness on particle transport from the target surface into the detector gas volume.
A simple Monte Carlo model in which fragments that escape from the target, but then collide with a different region of the target are removed is used.  
The efficiency of escape into the gas volume was found to have the form $1-\exp(-x/\gamma)$, where the parameter $\gamma$ represents the roughness.

The procedure used to generate the  $\cos(\theta)$ vs. energy distribution using Monte Carlo is as follows. 
A fission fragment is generated at a random depth in the actinide deposit and is propagated until stopping or escaping from the target, using the SRIM derived stopping power functions.  
There are a total of eight parameters that are varied in each Monte Carlo iteration and whose values are determined via a $\chi^2$ minimization with respect to the fissionTPC data.
The first parameter in the model is the thickness of the UF$_4$ deposit in the target.  
The next two parameters are $\beta$ and $\gamma$, which describe the target roughness.  The fourth parameter is the total fission kinetic energy.  
The fifth and sixth parameters represent an angle scatter after leaving the target, which is interpreted as the fragment scattering off of argon in the gas, i.e. being detected at angle different from that it was emitted.  A significant number of such tracks have been observed in the fissionTPC data.
These two parameters are the slope and intercept of the scattering angle as a function of fragment energy.  A fission anisotropy term is included to describe quantum anisotropy in the fission process.  Finally, an eighth term is included to represent the thickness of inert material on the surface of the target.

The data and the Monte Carlo model realization for the `best-fit' parameters that result from the $\chi^2$ minimization are shown for $^{238}$U and $^{235}$U  in Fig.~\ref{fig:EfficiencySimCompare}.  The slope towards lower energy at low $\cos(\theta)$ strongly constrains the target deposit thickness parameter.  
The variation in the intensity of the distribution as a function of $\cos(\theta)$ is most strongly influenced by the anisotropy term.  
The strong fall off in event statistics at low $\cos(\theta)$ is caused by preferential stopping of fragments in the target and the target roughness escape efficiency.  
The broadness of the distribution at low energy depends on the final scatter term in the gas and the surface roughness distribution.  
The $\chi^2$ minimization provides a single anisotropy term for each target, and an additional fitting procedure is needed to describe the change in quantum anisotropy as a function of energy.  
The ratio between the best-fit Monte Carlo model realization and data for each neutron energy bin is fit using a second-order Legendre polynomial. 

\begin{figure}[h]
\includegraphics[scale=0.43]{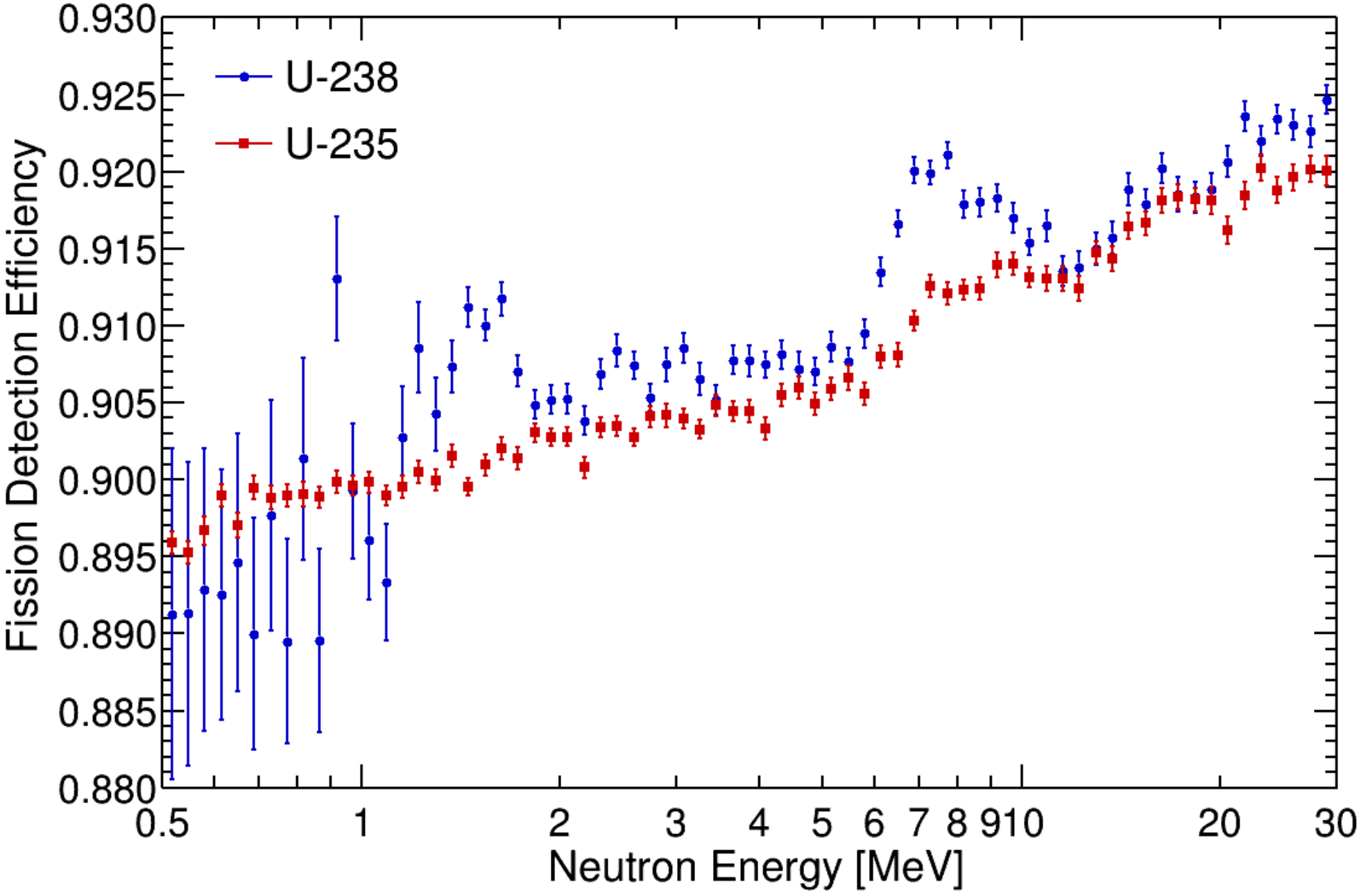}
\caption{\label{fig:Efficiency} Calculated fission detection efficiency for $^{238}$U and $^{235}$U.}
\end{figure}

\begin{table}
\caption{\label{tab:efficiency}Efficiency fit parameters and uncertainties for the two targets. }
\begin{ruledtabular}
\begin{tabular}{lcc}
Parameter & $^{238}$U & $^{235}$U \\
\hline
UF$_4$ thickness (mg/cm$^2$) & 0.346(1) & 0.292(1) \\
$\beta$ roughness & 0.00938(4) & 0.0101(1) \\
$\gamma$ roughness & 0.0267(7) & 0.0380(4) \\
Total energy (MeV) & 179.9(3) & 183.5(3) \\
Scatter offset (deg) & 20.3(1) & 21.3(1) \\
Scatter slope (deg/MeV) & -0.356(1) & -0.375(3) \\
Anisotropy & 1.201(5) & 0.935(3) \\
Inert thickness (mg/cm$^2$) & 0.0158(1) & 0.0156(1) \\
\end{tabular}
\end{ruledtabular}
\end{table}

The fragment transport and anisotropy best-fit parameters are combined to calculate the fission fragment detection efficiency as a function of energy, and Monte Carlo error propagation (see Section \ref{sec:ErrorPropagation}) is used to calculate the efficiency uncertainty from the fitting procedure covariances (Fig. \ref{fig:Efficiency}). 
The Monte-Carlo transport and anisotropy model, using the best-fit parameters directly describe the fraction of fission fragments that enter the active volume of the fissionTPC and that would pass analysis selection cuts.
The upward slope as a function of energy is a consequence of the kinematic boost from neutron momentum transfer.  
The fission fragment detection volume for both targets is downstream from the neutron beam, and the momentum transfer increases the number of fragments entering that volume.  
The energy-dependent structure in the efficiencies result from the quantum anisotropy of fission, which must be measured for each energy bin.  
The larger uncertainties at low energy for $^{238}$U are due to low statistics below the fission threshold.

\begin{figure*}[t!]
\includegraphics[scale=0.8]{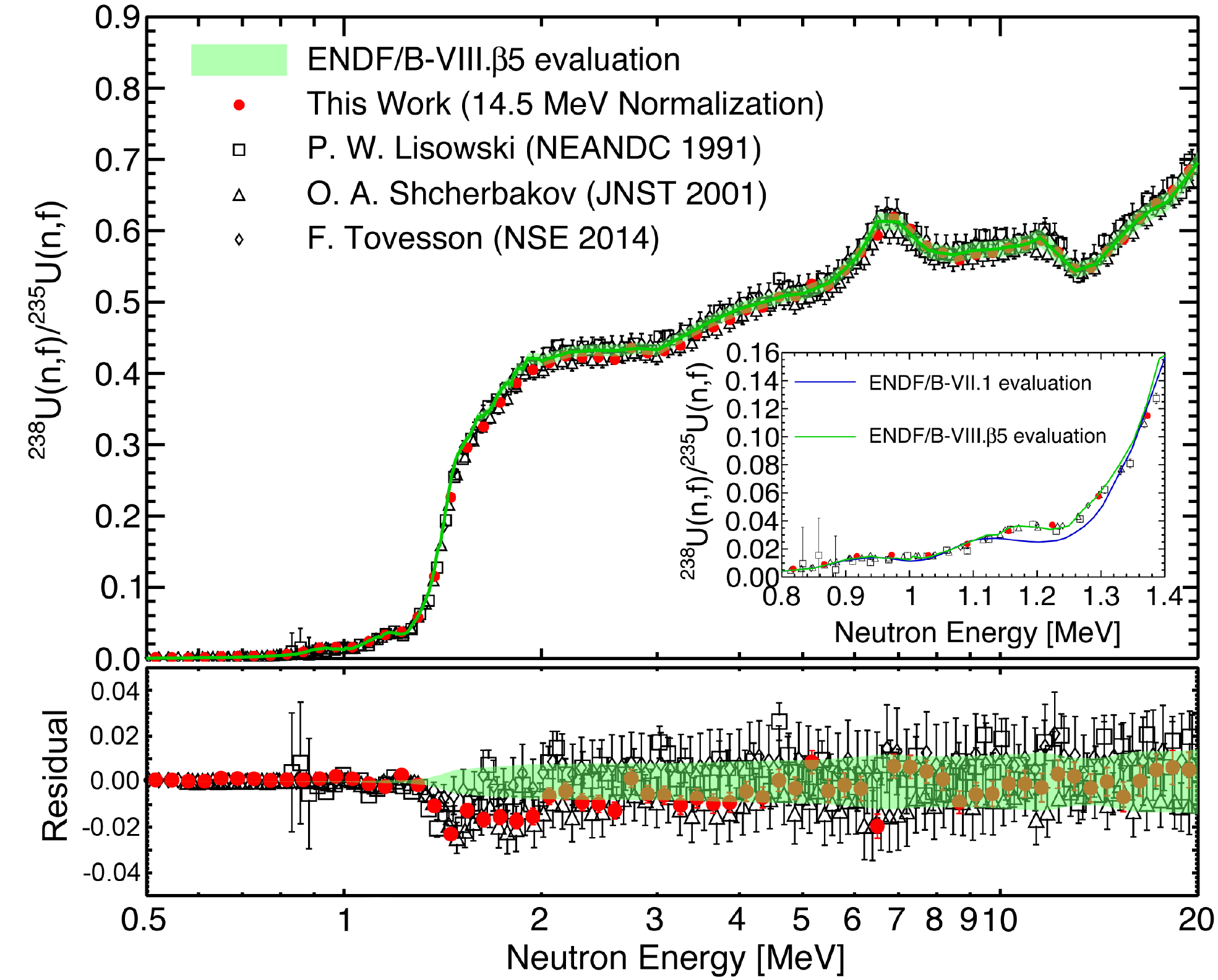}
\caption{\label{fig:DataCompareFull} The normalized $^{238}$U(n,f)/$^{235}$U(n,f) cross section ratio measured in this work, compared with the three most recent measurements and the ENDF/B-VIII.$\beta$5~\cite{ENDFB8B5} evaluation shown with the evaluated uncertainty.  In many cases, the uncertainty on the measured data is smaller than the symbol.  The inset shows a comparison to the ENDF/B-VII.1~\cite{Chadwick2011NDS}
 and ENDF/B-VIII.$\beta$5 evaluations near $1.2$~MeV.  The lower plot shows the residual of the four data sets and ENDF/B-VIII.$\beta$5, shown with the evaluated uncertainty.}
\end{figure*}

The best-fit parameters and uncertainties for the two targets are shown in Table \ref{tab:efficiency}.  While the model was based upon a physical description of the processes affecting the efficiency, the parameters involved are not necessarily physically precise values, as a number of compensating effects can occur.  For example, uncertainties in the SRIM-derived stopping powers would be correlated with the target thickness, the total energy would correlate with the choice of digitizer channel to energy conversion factor, and the anisotropy would correlate with the choice of TPC drift velocity.  The purpose of these calculations is to describe the fission fragment distribution as accurately as possible, and extrapolate the data below an energy threshold using the efficiency model.  Compensating factors like these should not significantly affect the extrapolation.

\subsection{\label{sec:ErrorPropagation}Uncertainty Propagation}

The ratio defined in Eq.~\ref{eq:CSRatio} can be described by a probability distribution for each neutron energy range, and the covariance of these values is calculated via Monte Carlo error propagation.  
Each term in the ratio has an assigned uncertainty, and some terms have fit parameter covariance matrices.  
The product of the transposed Cholesky decomposition~\cite{Gentle1998Book} and random Gaussian vectors are used to generate  $100$~realizations of all ratio terms.  The same Gaussian vectors are applied across the full neutron energy range, and a ratio covariance as a function of energy can be found by analyzing the ratio distributions for pairs of energies.  The $C_{bb}$ and $\epsilon_{ff}$ terms use full covariance matrix error propagation, the $G$ term is considered fully correlated, and $C_{ff}$ is considered uncorrelated as a function of energy.  
The anisotropy contribution was solved for independently of the main efficiency model covariance matrix, i.e. the parameters describing anisotropy were varied around their best-fit values independently of those for the efficiency model.

As mentioned in Section \ref{sec:FissionCuts}, the PID energy cut is varied over a range of values, where the minimum value is above $\alpha$-particle and recoil contaminants, and the maximum value removes a small fraction of fission events in the fission distribution.  The fission fragment detection efficiency in the cross section ratio 
is calculated for each cut variation, and any dependence of the ratio on the cut energy is considered a residual efficiency uncertainty.  For this analysis, the cut is varied~100 times across a uniform distribution.

\begin{figure}[h]
\includegraphics[scale=0.43]{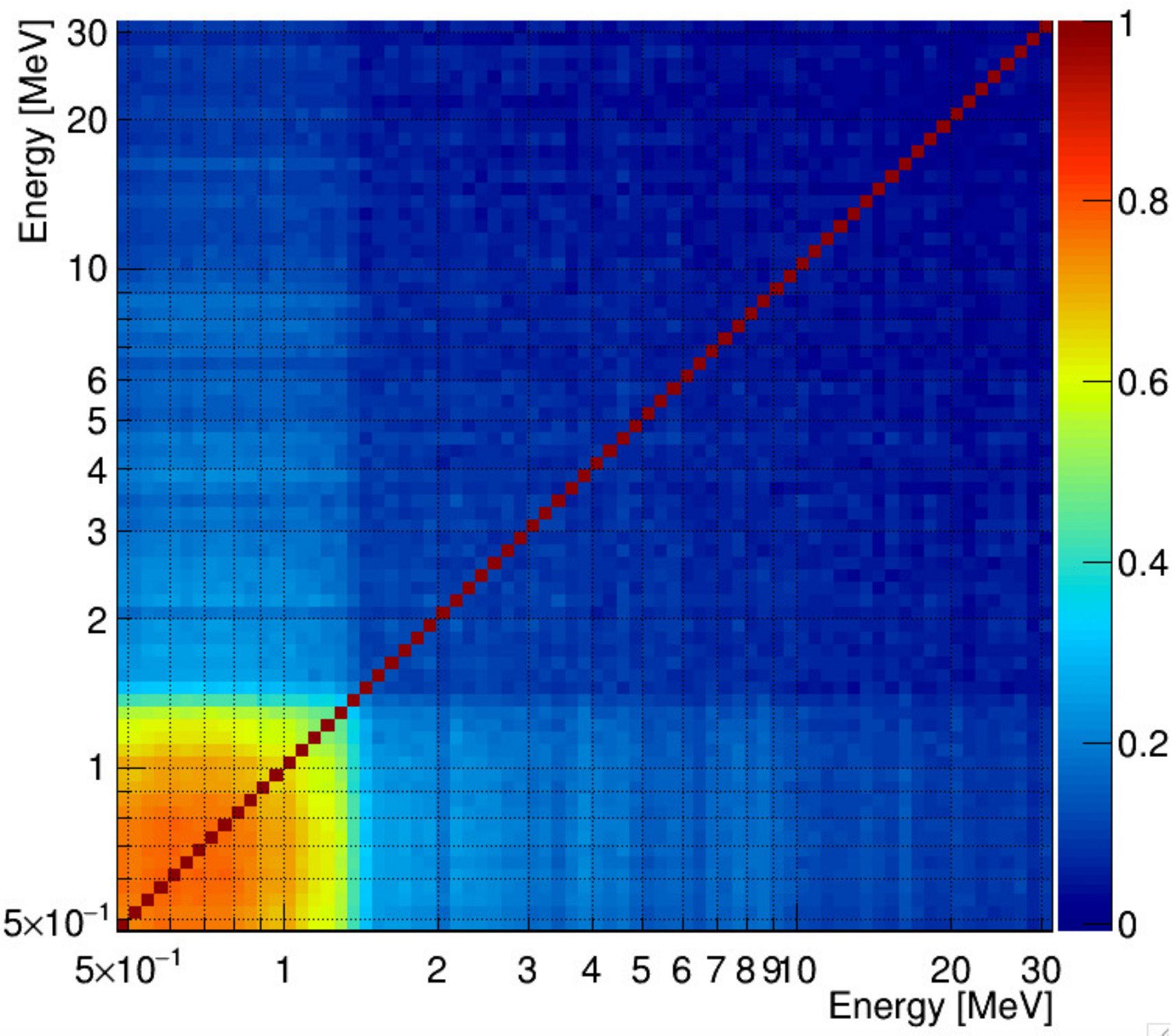}
\caption{\label{fig:U8U5Correlation} The $^{238}$U(n,f)/$^{235}$U(n,f) correlation matrix measured in this work.  At low neutron energy, the contaminant correction becomes the largest source of uncertainty, resulting in a large correlated region in the correlation matrix.  The contaminant correction is a fixed value at all energies and as the ratio becomes small at low energy, a large relative uncertainty results.  The z-axis represents the value of the correlation matrix elements.}
\end{figure}

The final cross section ratio is calculated by performing the 100 Monte Carlo term variations for each of 100 energy cut variations, resulting in 10,000 values in the cross section ratio distribution for each energy bin.  The mean of the ratio distribution is calculated for each energy bin, and covariance is calculated with pairs of energy bins.  The normalized cross section ratio with all uncertainties is shown in Fig.~\ref{fig:DataCompareFull}, compared to the ENDF/B-VII.1~\cite{Chadwick2011NDS} and ENDF/B-VIII.$\beta$5~\cite{ENDFB8B5} evaluations.  The correlation matrix for the cross section ratio is shown in Fig. \ref{fig:U8U5Correlation}, where the z-axis represents the value of the correlation matrix elements.  The ratio is normalized to the ENDF/B-VIII.$\beta$5 evaluation at $14.5$~MeV.  The covariance matrix is related to the correlation matrix by the uncertainties shown in Fig.~\ref{fig:DataCompareFull}.

\section{\label{sec:Discussion}Discussion}

\begin{figure}[h]
\includegraphics[scale=0.43]{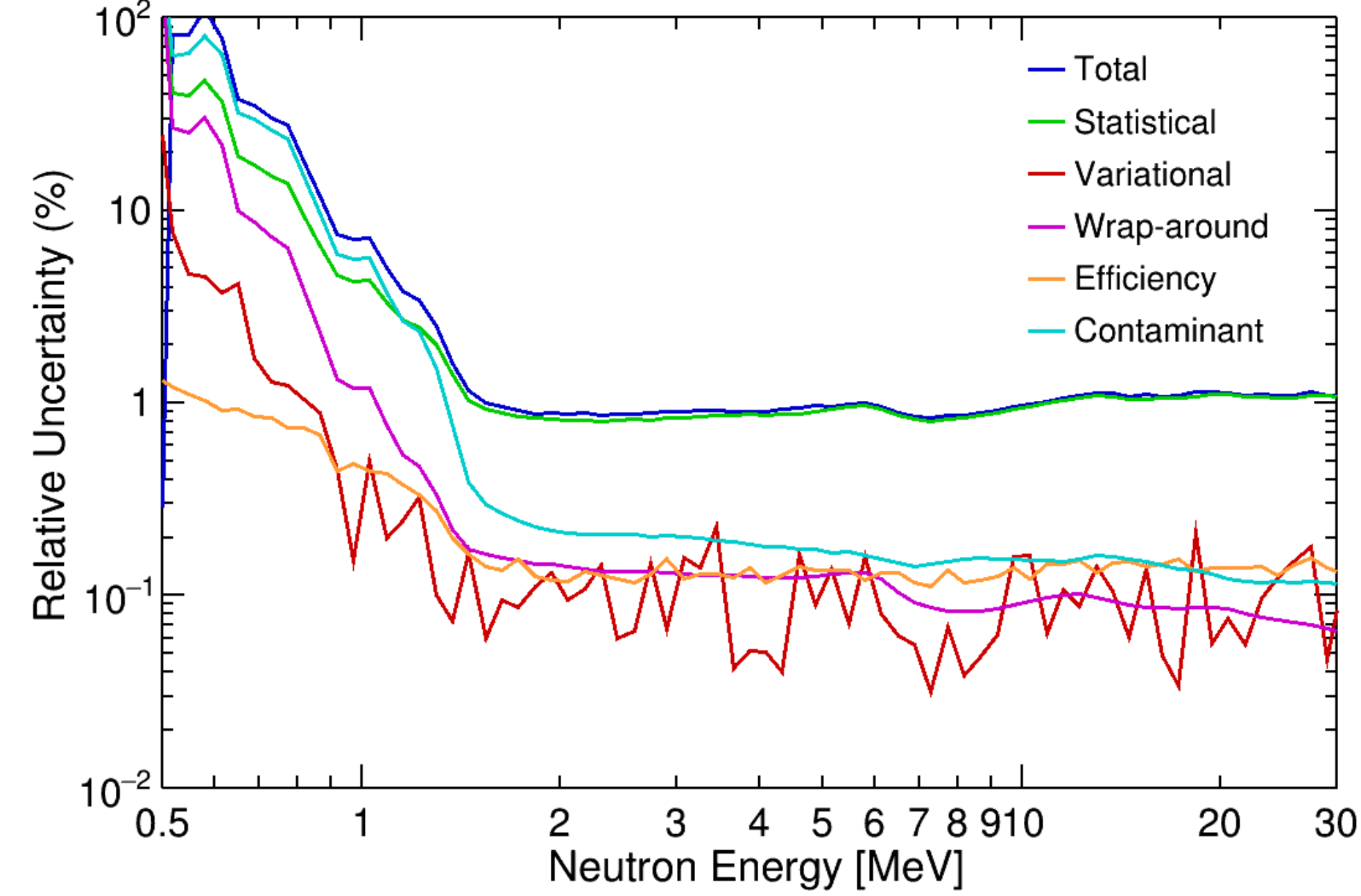}
\caption{\label{fig:Uncertainty} Uncertainty contributions to the $^{238}$U(n,f)/$^{235}$U(n,f) cross section ratio.  At low neutron energy, the contaminant correction becomes the largest source of uncertainty, and statistical uncertainty is largest at high energy.  The contaminant correction is a fixed value at all energies and as the ratio becomes small at low energy, a large relative uncertainty is found.}
\end{figure}

The various uncertainty contributions to the measured cross section ratio can be isolated by enabling individual contributions in the error propagation procedure (Fig. \ref{fig:Uncertainty}).  The largest contribution to the total uncertainty at high energies is the statistical uncertainty, while at low energies the contaminant uncertainty dominates, because the cross section ratio drops dramatically below the fission threshold.  The efficiency fit contributes the next largest uncertainty at high energy, although the contribution is significantly smaller than the statistical uncertainty.  The residual uncertainty refers to the sensitivity of the cross section ratio to variations in the energy-based PID cut, which is similar to the efficiency uncertainty at higher energy.  The wrap-around correction is a minor contribution to the total uncertainty.

The cross section ratio has been normalized to the ENDF/B-VIII.$\beta$5 evaluation at 14.5 MeV, as the uncertainty at this energy is relatively small~\cite{ENDFB8B5}.  The beam flux $\Phi$ and actinide density $N$ factor out of Eq. \ref{eq:CSRatio} when normalizing, which removes the uncertainty associated with those terms.  The neutron beam flux was calculated with the measured proton distribution in the fissionTPC, resulting from neutrons scattering off of hydrogen in the drift gas, and it was found that a small tilt in the detector or gain variations across the pad plane could result in a difference between the measured proton distribution and neutron flux at the target.

With thick-backed actinide targets which overlap in the $x$-$y$ dimensions, the fission and $\alpha$-particle spatial distribution can be used as a second method for calculating the neutron flux, and this would not be sensitive to the tilt of the detector or gain variations.  The target used for this measurement has two half-disk actinide deposits on a thin carbon-backed target which did not have any actinide overlap in $x$ and $y$, and such a correction could not be made.  Future measurements will include thick-backed targets with actinide deposits on both sides.

\begin{figure}[h]
\includegraphics[scale=0.43]{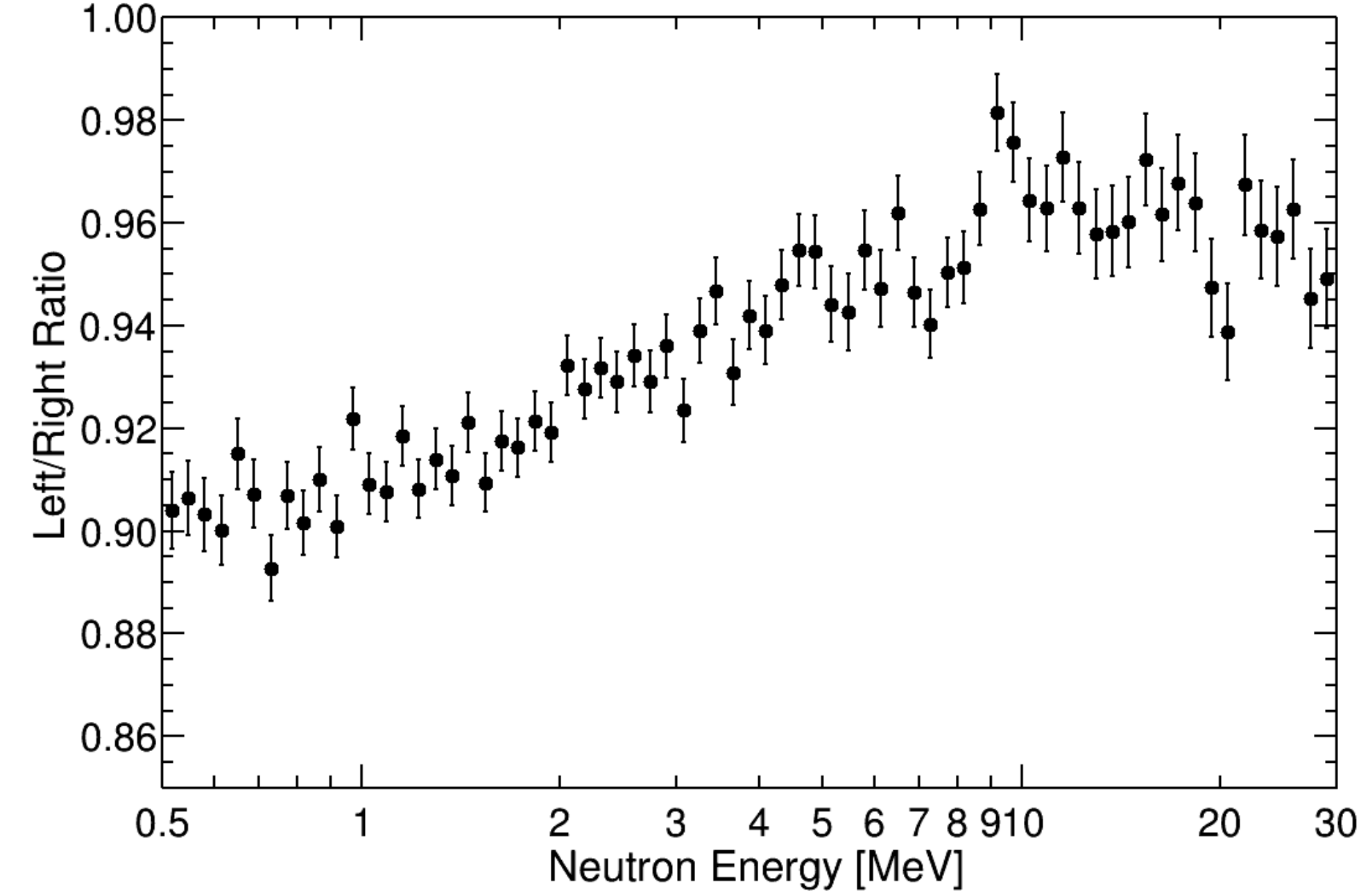}
\caption{\label{fig:DataCompareZoom} The fission ratio of the left and right half of the $^{235}$U target measured from this work.} 
\end{figure}

Typical neutron-induced fission cross section measurements have stacks of targets that have roughly the same spatial distribution of actinide deposits and neutron flux.  
The ability of the fissionTPC to identify energy, length, track angle, and start position allows for the thin-backed half-disk target used in this work.  
It was previously assumed that the neutron beam flux varied spatially, but that the neutron energy spectrum did not.
To test this, a ratio of fission counts was taken between different regions of the target, and a 7\% variation in neutron flux as a function of energy was observed.  
This ratio can be seen in Fig.~\ref{fig:DataCompareZoom}, with a gradual increase occurring between $0.5$ and $10$~MeV.

MCNP simulations \cite{Pelowitz2011LA} show that this is due to an intervening neutron collimator exposing off-axis areas of the fission foil to different sections of the tungsten spallation target. 
As the proton beam slows down in the tungsten, the neutron spectrum softens leading to a spatially varying neutron energy spectrum.
Such a flux variation should only be observed in the direction of the beam, which is parallel to the ground.  The half-disk targets used in this measurement are bisected by a plane consistent with the beam direction, and therefore flux variations should not be observed between the two targets.  
This was confirmed experimentally in a separate measurement of different actinides, which had deposits rotated 90$^{\circ}$ relative to the target used in this experiment.  In that confirmation measurement, the top/bottom ratio was consistent with unity.

\begin{figure}[h]
\includegraphics[scale=0.43]{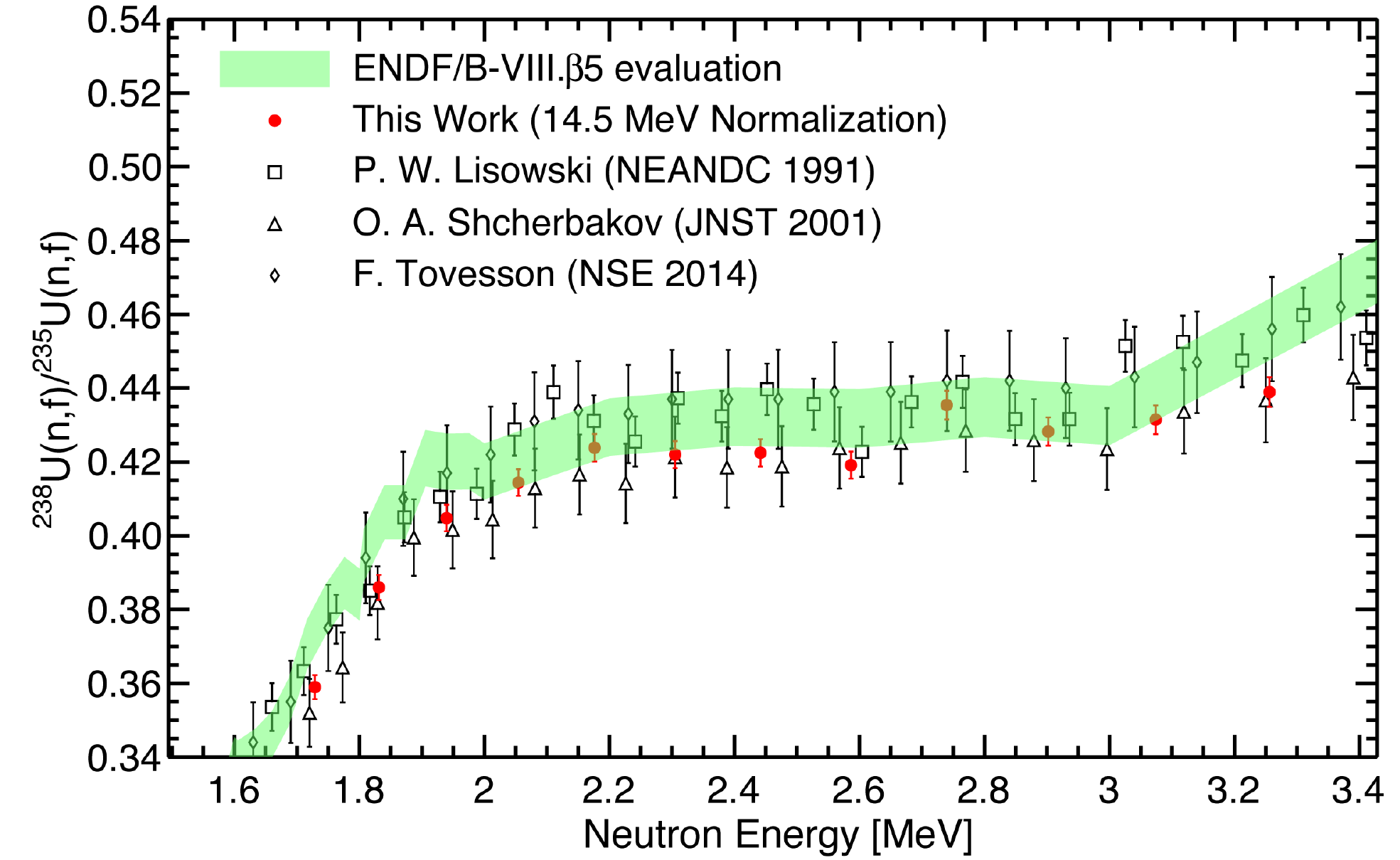}
\caption{\label{fig:DataCompareMid} The normalized $^{238}$U(n,f)/$^{235}$U(n,f) cross section ratio measured in this work, compared with the three most recent measurements and to the ENDF/B-VIII.$\beta$5 \cite{ENDFB8B5} evaluation shown with the evaluated uncertainty.  The neutron energy range $1.6$ to $3.4$~MeV is shown, and the new data is seen to differ from ENDF/B-VIII.$\beta$5 by $\sim$2.5\% at $2.4$~MeV.}
\end{figure}

The result here is compared to the three most recent $^{238}$U(n,f)/$^{235}$U(n,f) measurements, as well as to the ENDF/B-VIII.$\beta$5  evaluation in Fig.~\ref{fig:DataCompareFull}.
There was a recent change in  ENDF/B-VIII.$\beta$5  for the $^{238}$U(n,f) cross section, which resulted from a 40\% change in the evaluation at $1.2$~MeV.
 A comparison of this work to ENDF/B-VII.1~\cite{Chadwick2011NDS} and ENDF/B-VIII.$\beta$5~\cite{ENDFB8B5} is shown with three previous data sets in the inset of Fig. \ref{fig:DataCompareFull}.  
 The $^{238}$U(n,f)/$^{235}$U(n,f) cross section ratio measured in this work agrees with most recent data, and provides support for the recent change in the evaluation.

A significant difference in the cross section is observed between this work and past measurements in the energy range $2-3$~MeV (Fig.~\ref{fig:DataCompareMid}), with this work most closely agreeing with Shcherbakov~\cite{Shcherbakov2002JNST}. 
The disagreement between this measurement and  ENDF/B-VIII.$\beta$5  is greatest ($\sim$2.5\%) near $2.4$~MeV.  
The cross section ratio presented here is normalized to ENDF/B-VIII.$\beta$5 at $14.5$~MeV neutron energy, and the disagreement at $2.4$~MeV indicates a difference in the cross section ratio shape.  
Without an absolute normalization, we are not able to determine the energy range in which the disagreement occurs.

\section{\label{sec:Conclusions}Conclusions}

The normalized $^{238}$U(n,f)/$^{235}$U(n,f) cross section ratio has been measured using the fissionTPC over the neutron energy range $0.5$ to $30$~MeV.  The fissionTPC allowed for a detailed analysis of systematic uncertainties by providing particle information which is unique to this technique.
By fitting the distributions of fission fragment energy and angle using a Monte Carlo simulation of the target, an efficiency correction factor could be applied to the measured fission event count, which allows for a higher energy cut to exclude $\alpha$-particle and neutron recoil backgrounds.
Error propagation of the wrap-around and efficiency fits were combined with a variational analysis to produce an accurate measure of the systematic covariance for the cross section ratio.

The cross section ratio presented here is normalized to the ENDF/B-VIII.$\beta$5 evaluation at 14.5 MeV.
This allows the shape of the ratio to be reported over the full neutron energy range without the large neutron beam flux uncertainty introduced by the target geometry.  Future measurements will be performed with thick-backed targets and back-to-back actinide deposits which will allow for precise determination of the neutron beam flux and absolute normalization.

This cross section ratio has the potential to impact other measurements, because the $^{238}$U(n,f) cross section is a standard used in neutron flux measurements, and can cause correlations between different nuclear data sets.  This new data
provides additional support for the recent 40\% change of the $^{238}$U(n,f) cross section reflected in the ENDF/B-VIII.$\beta$5 evaluation.  In addition, the measured cross section ratio shape can be used to improve nuclear physics knowledge of the compound nuclei by fitting the data with nuclear reaction models.

\begin{acknowledgments}
The authors would like to thank Caleb Mattoon for valuable discussions. This work performed under the auspices of the U.S. Department of Energy by Lawrence Livermore National Laboratory under Contract DE-AC52-07NA27344.  The neutron beam for this work was provided by LANSCE, which is funded by the U.S. Department of Energy and operated by Los Alamos National Security, LLC, under contract DE-AC52-06NA25396.    University collaborators acknowledge support for this work from the U.S. Department of Energy Nuclear Energy Research Initiative Project Number 08-014, the U.S. Department of Energy Idaho National Laboratory operated by Battle Energy Alliance under contract number 00043028-00116, the DOE-NNSA Stewardship Science Academic Alliances Program, under Award Number DE-NA0002921, and through subcontracts from LLNL and LANL.
\end{acknowledgments}

\bibliography{u8u5ratio}

\end{document}